\documentclass[sigconf]{acmart}
\renewcommand\footnotetextcopyrightpermission[1]{} 
\setcopyright{none} 

\usepackage{pgfplots}
\usepackage{array}

\begin{document}
\title{Detecting Cyberattacks in Industrial Control Systems Using Convolutional Neural Networks}

\author{Moshe Kravchik, Asaf Shabtai}
\affiliation{%
  \institution{Department of software and information system engineering \\Ben Gurion university, Be'er Sheba, Israel}
}
\email{moshekr@post.bgu.ac.il, shabtaia@bgu.ac.il}

\begin{abstract}
This paper presents a study on detecting cyberattacks on industrial control systems (ICS) using unsupervised deep neural networks, specifically, convolutional neural networks.
The study was performed on a Secure Water
Treatment testbed (SWaT) dataset, which represents a scaled-down version of a real-world industrial water treatment plant.
e suggest a method for anomaly detection based on measuring the statistical deviation of the predicted value from the observed value.
We applied the proposed method by using a variety of deep neural networks architectures including different variants of convolutional and recurrent networks. 
The test dataset from SWaT included 36 different cyberattacks. 
The proposed method successfully detects the vast majority of the attacks with a low false positive rate thus improving on previous works based on this data set. 
The results of the study show that 1D convolutional networks can be successfully applied to anomaly detection in industrial control systems and outperform more complex recurrent networks while being much smaller and faster to train.
\end{abstract}



\maketitle

\section{INTRODUCTION}
Industrial control systems (ICS) are widely used and vital to the operation of various sectors,  including the pharmaceutical industry, manufacturing, and critical infrastructures, such as electricity, water treatment plants, and oil refineries. 
Historically, these systems ran on proprietary hardware and software in physically secure locations, but more recently they have adopted common information technology (IT) technologies and remote connectivity.
These changes increase the likelihood of cyber security vulnerabilities and incidents \cite{Stouffer:2011:SGI:2206293}.
A number of high impact cyber attacks were reported in recent years, including the attack on a power plant in the Ukraine in December 2015 \cite{web:Ukrain_power_hack}, the infamous Stuxnet malware that targeted nuclear centrifuges in Iran, and recent attacks on a Saudi oil company \cite{web:Saudi_Aramco_hack}.

Thus, the ability to detect cyberattacks on ICSs has become a critical task. 
One of the approaches used to address this problem involves utilizing traditional IT network-based intrusion detection Systems (IDSs) to identify malicious activity. 
In this work we focus on an anomaly detection approach, in which we attempt to detect anomalous behavior of the system on the physical level. 
This approach is based on the assumption that the ultimate goal of the attacker is to influence the physical behavior of the system, and aims at protecting the system beyond the network-level line of defense. 
Physical level-based anomaly detection can also facilitate early detection and remediation of faulty equipment, which can be of a great economical value.

Anomaly detection methods can be based on rules or models of the system \cite{pasqualetti2011cyber} \cite{teixeira2012attack} \cite{jones2014anomaly}. 
Unfortunately, creating a precise model of complex physical processes is a very challenging task. 
It requires an in depth understanding of the system and its implementation, which is a time consuming and cannot scale up to large and complex systems. 
Another approach that has recently been the focus of interest utilizes machine learning to model ICSs and detect anomalous behavior. 
A number of works using supervised machine learning for anomaly detection in ICSs have been published recently \cite{beaver2013evaluation} \cite{hink2014machine}.
This approach requires labeled training data for normal and attack scenarios, however, labeled data for cyberattacks may be difficult to acquire, and this data will naturally not include unknown attack classes.  
Recently, unsupervised machine learning was shown to be effective \cite{goh2017anomaly} \cite{inoue2017anomaly} for detecting cyberattacks using data obtained from a dedicated water plant testbed  (SWaT)\cite{goh2016dataset} that was built to support research related to the design of secure cyber-physical systems (CPSs). 
In \cite{goh2017anomaly} the authors use recurrent neural network (RNN) to detect attacks on a single stage of a six-stage water purification process, and report detection of nine out of ten attacks with four false positives. 
In addition, \cite{inoue2017anomaly} compare the performance of a long short term memory (LSTM)-based deep neural network (DNN) and one-class SVM to detect attacks on all stages of the same process.
They show some improvement, with $F1 = 0.8$ for DNN when measured per log record. 
However, when relating to specific attacks, 23 out of 36 have a recall = 0.

In this work, we present further research on different architectures of unsupervised DNN to detect cyberattacks on all stages of the SWaT dataset. 
The contributions of this papers are:
\begin{itemize}
\item a method for anomaly detection for multi-variant industrial process time series data;
\item successful use of 1D convolutional neural networks (CNN) for detecting anomalies and cyberattacks in ICS data with few false positives; the model is able to achieve this performance by combining the detection for individual stages in the process;
\item a comparison of the efficiency of different neural network architectures for anomaly detection in ICSs.
\end{itemize}


\section{RELATED WORK}
Anomaly and intrusion detection in industrial control systems (also called cyber physical systems) have been extensively studied.
A number of comprehensive surveys are dedicated to the classification of techniques and methodologies in this area (e.g., \cite{han2014intrusion} and \cite{mitchell2014survey}).   
A well-known approach to intrusion detection in ICS is based on modeling and simulation of the system \cite{pasqualetti2011cyber}, \cite{teixeira2012attack}. 
Practical problems with this approach are the need for precise knowledge of the system's design and configurations, as well as the need to accurately modeling the system's complex physical behavior.
According to Mitchell \textit{et al.} \cite{mitchell2014survey}, ICS anomaly detection methods include knowledge and behavior-based methods.
Knowledge-based detection techniques search for known attack characteristics, similar to malware signature techniques in IT intrusion detection.
While having low false positive rates, these approaches require maintaining an updated dictionary of attack signatures and are ineffective against zero-day attacks.
In contrast, behavior-based techniques search for anomalies in runtime behavior. 
These techniques are more common in ICS intrusion detection, since ICS systems are automated and present more regularity and predictability than typical IT systems.
The proposed method utilizes behavior-based techniques.

Another approach to classifying intrusion detection methodologies is based on the data being monitored.
Numerous studies have presented network traffic-based intrusion detection \cite{linda2009neural}. 
Ghaeini \textit{et al.} \cite{ghaeini2016hamids} use this approach on the SWaT dataset used in our study.
In this work we have study an alternative approach based on the data collected from the sensors and actuators, thus focusing on the system behavior at the physical layer.

Anomaly detection techniques used in ICS intrusion detection can be broadly divided into supervised, unsupervised and semi-supervised techniques.
Supervised techniques require prior labeling of the system behavior, including the samples of malicious behavior. 
Acquiring precise and representative labeled data is very hard to obtain in practice, and this data is highly dependent on the specific system. 
Therefore, most of recent research on ICS intrusion detection uses unsupervised (unlabeled data from real data) \cite{hinton1999unsupervised} and semi-supervised (training from a set of clean data with no anomalies) approaches.
Unsupervised SCADA intrusion detection was investigated in \cite{maglaras2016novel} which describes a technique based on one-class SVM and k-means clustering.
Semi-supervised learning approaches are trained using a collection of "good" data, which is assumed to completely represent normal system behavior and contain no attacks. 
While both assumptions should be examined closely, they can be fulfilled in many practical situations.
Semi-supervised methods usually have lower false-positive rates than their fully unsupervised alternatives \cite{zhang2011distributed}, \cite{zhang2011artificial}.

Multiple machine learning techniques are used in ICS anomaly detection. 
They include support-vector machine \cite{maglaras2016novel} \cite{hink2014machine}, random forest \cite{beaver2013evaluation} as well as artificial neural network \cite{linda2009neural}, \cite{gao2010scada} and others.
In the past few years some research has been performed applying deep and recurrent neural networks to this area \cite{malhotra2015long}, \cite{goh2017anomaly}, \cite{inoue2017anomaly}.
In our research 1D convolutional neural networks (1D CNNs) are used, and demonstrate superior attack detection abilities and higher $F1$ scores than previously published papers \cite{goh2017anomaly}, \cite{inoue2017anomaly}.
Previously, 1D CNNs were used for detecting faulty motor bearings based on univariant motor current data \cite{ince2016real}.
In this study we apply 1D CNNs to multivariant time series data and use it to detect multiple instances of cyberattacks.
To the best of our knowledge, our work is the first to use 1D CNNs for cyberattacks detection in ICSs.

\section{SECURE WATER TREATMENT (SWaT) TESTBED DATASET}
The Secure Water Treatment (SWaT) testbed was built by the Singapore University of Technology and Design in order to provide researchers with data collected from a realistic complex ICS environment. 
Although a detailed description of the testbed and the dataset can be found in \cite{goh2016dataset}, we will provide a brief description below.
The testbed is a fully operational scaled down water treatment plant that produces purified water.

\begin{figure}[b!]
\centering
\includegraphics[clip, trim=6cm 12cm 6cm 12cm]{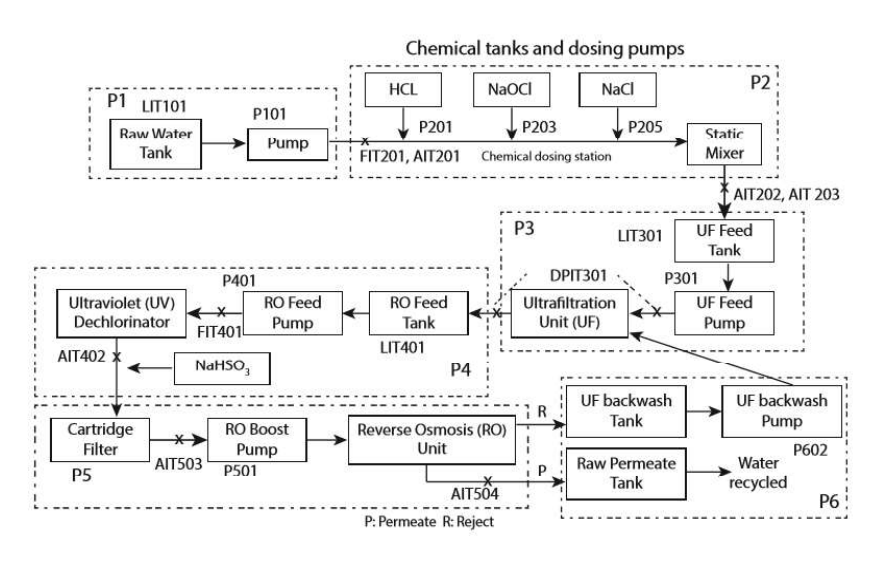}
\caption{SWaT Testbed Process Overview \cite{goh2016dataset}}
\label{fig:swat}
\end{figure}

\begin{figure}[b!]
\centering
\includegraphics[clip, scale = 0.3, trim=0cm 9cm 0cm 9cm]{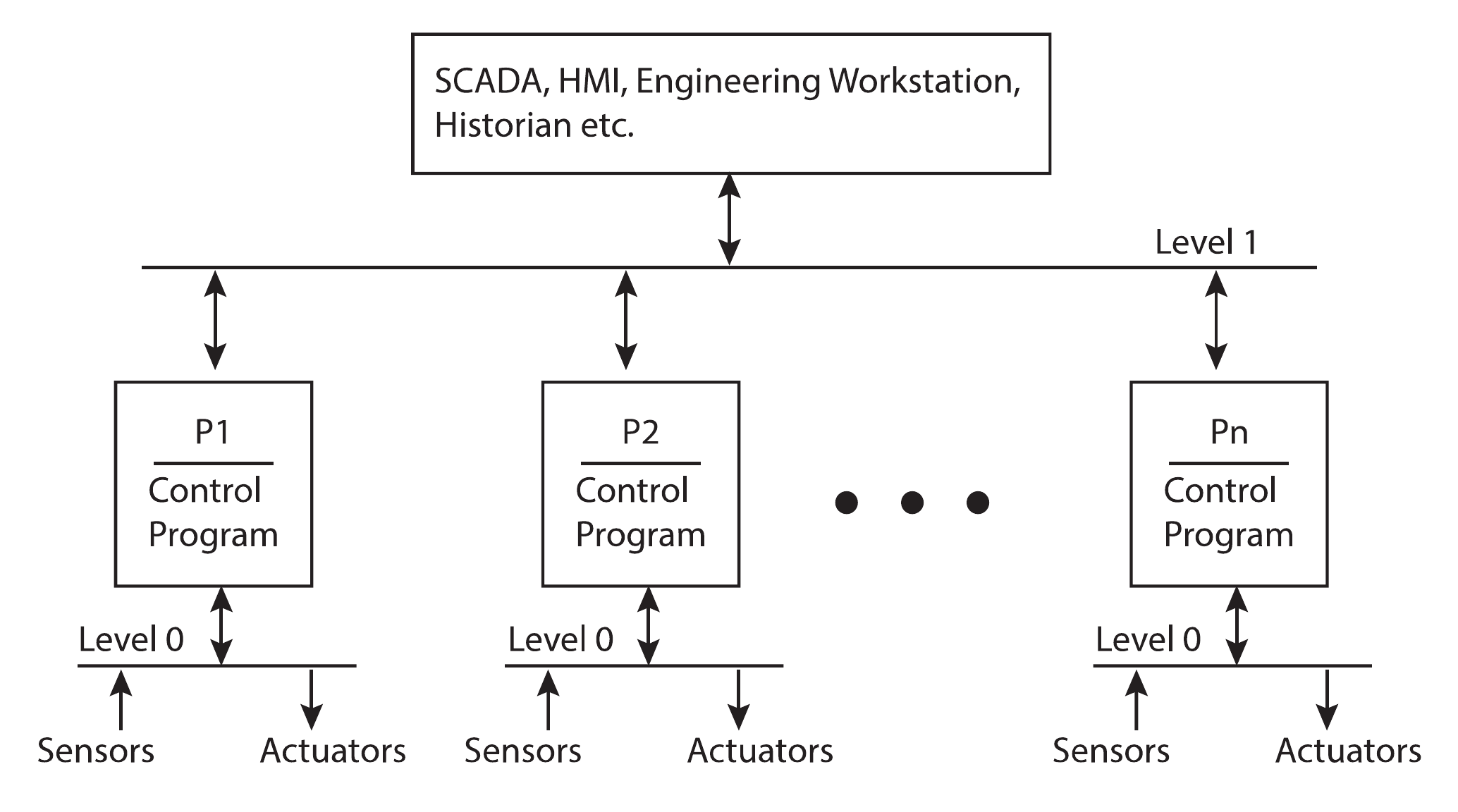}
\caption{SWaT Testbed Network Overview \cite{goh2016dataset}}
\label{fig:swat_nw}
\end{figure}

As shown in Figure \ref{fig:swat}, the water goes through a six-stage process, each stage (from P1 to P6) is equipped with a number of sensors and actuators.
Sensors include flow meters, water level meters, conductivity and pH analyzers and more.
Actuators include pumps that transfer water from stage to stage, pumps that dose chemicals, and valves that control inflow.
The process is not circular, and the water from P6 is disposed.
The sensors and the actuators at each stage are connected to the corresponding PLC (programmable logic controller), and the PLCs are connected to the SCADA system as shown in Figure \ref{fig:swat_nw}.

The data from all of the sensors and actuators was logged every second in a Historian server, and this data was used for training and testing the system.
The dataset also contains recorded network traffic, but this was not used in our work.
The dataset contains seven days of recording under normal conditions and an additional four days of recording when 36 attacks were conducted.
The attacks were conducted by altering the network traffic in the Level 1 network, spoofing the sensors values and issuing fake SCADA commands.
Attacks include (a) attacks that target a single stage of the process and (b) attacks acting simultaneously at different stages.
A table listing the attacks and the corresponding attack times, attack points, and expected and factual outcomes is provided \cite{goh2016dataset}.
For example, attack 21 aims at overflowing the tank at the stage P1.
For that purpose, the value of the level sensor LIT101 is fixed at 700mm while the motorized valve controlling water inflow is kept open for 12 minutes.

\begin{figure}[t!]
\centering
\includegraphics[clip, scale = 0.4, trim=0cm 0cm 0cm 0cm]{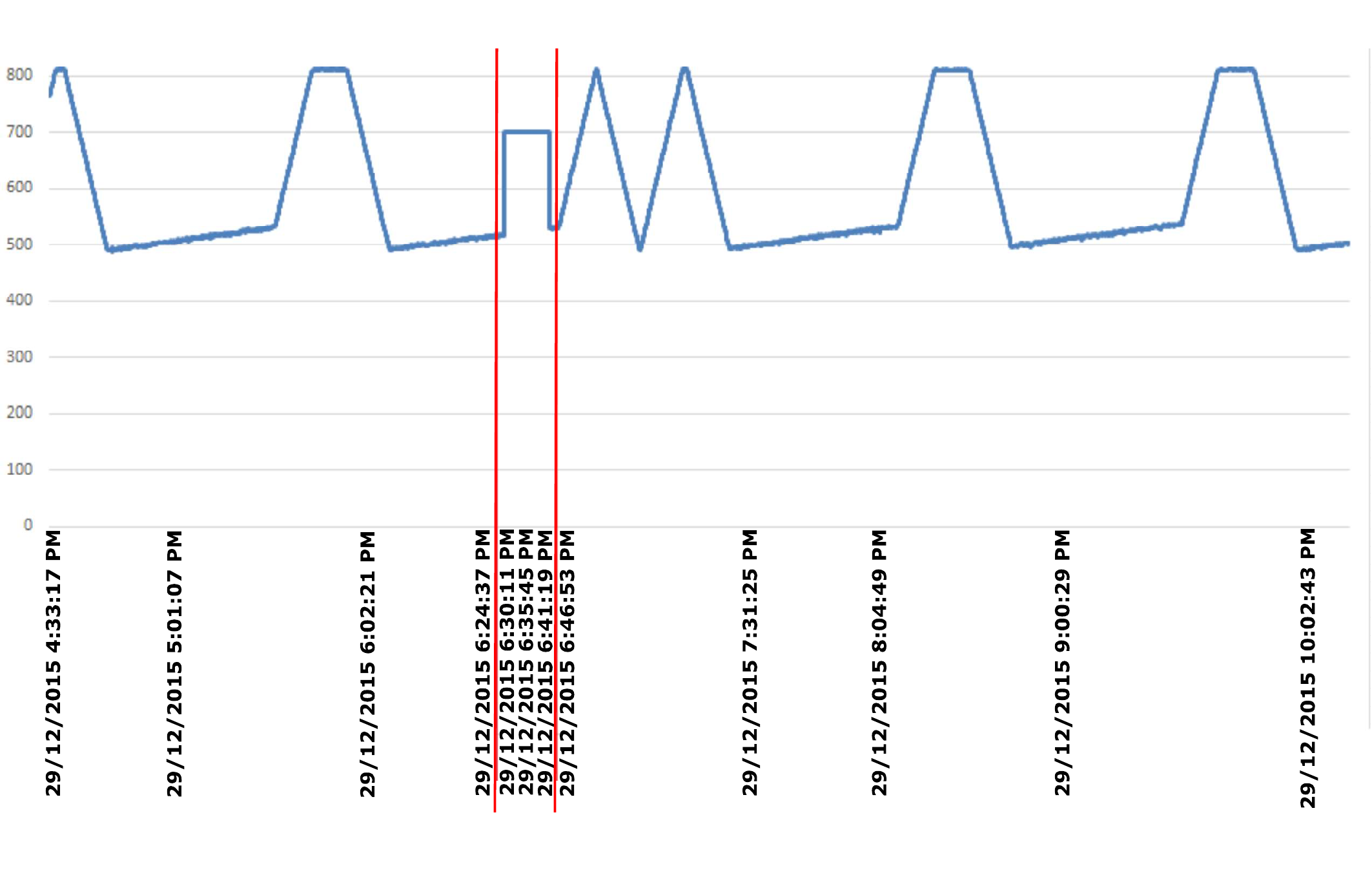}
\caption{Attack 21 on the LIT101 sensor \cite{goh2016dataset}}
\label{fig:swat_att_21}
\end{figure}

Figure \ref{fig:swat_att_21} (from \cite{goh2016dataset}) shows the attack, its influence on the LIT101 sensor and the time it takes the system to stabilize.
The entire dataset contains 946,722 records, labeled as either attack or normal, with 51 attributes corresponding to the data from sensors and actuators.

\section{BACKGROUND ON NEURAL NETWORKS}

\subsection{Recurrent Neural Networks}
This section provides a brief introduction to recurrent neural networks (RNNs), for a detailed discussion, please refer to Lipton \textit{et al.} \cite{lipton2015critical}.
Traditional feed-forward neural networks assume independence among samples, resetting the state after each one.
This approach has shown strong results for many tasks, but has limitation when the samples are related in time, such as audio signals, speech, video and many other important areas.
The main difference between RNNs and the standard feed-forward neural networks is the ability to maintain the state between the inputs and selectively pass information based on the internal state.
While very powerful, RNNs were found to be hard to train due to a vanishing and exploding gradient problem caused by back-propagations across many time steps. 
This problem has been addressed by a special type of RNNs, Long Short Term Memory (LSTM) networks \cite{hochreiter1997long} as well as the use of the Truncated Back-propagation Through Time (TBTT) technique \cite{williams1989learning}. 
We employed both techniques in our work.


\subsection{Convolutional Neural Networks}
Convolutional neural networks (CNN) are feed-forward neural networks that became popular in image processing after the groundbreaking work of LeCunn \textit{et al.} \cite{lecun1990handwritten}.
CNNs significantly increase the efficiency of neural networks by applying \textit{convolutions}, which are basically filters, to small regions of the input instead of performing matrix multiplication over the entire image at once.


While traditional CNNs used in image processing are 2D, 1D CNNs can be successfully used for time series processing, because time series have a strong 1D (time) locality that can be successfully extracted by convolutions \cite{lecun1995convolutional}.
In this work we show that 1D CNNs can be successfully used for detecting cyberattacks in complex multivariant ICS data.

\section{ANOMALY DETECTION METHOD}\label{ssec:anom_detect_method}
We used the following statistical window-based anomaly detection method in our research.
A neural network model predicts the future values of the data features based on previous values. 
We provide the network with a sequence $(x_0, x_1, \ldots, x_{n-1})$ to predict a sequence $(y_n, y_{n+1}, \ldots, y_{n+m})$ where  $x_i$ and $y_i$ are input and output vectors of feature values at time $i$.

\begin{figure}[t!]
\centering
\includegraphics[clip, trim=0cm 0cm 0cm 0cm, scale=0.4]{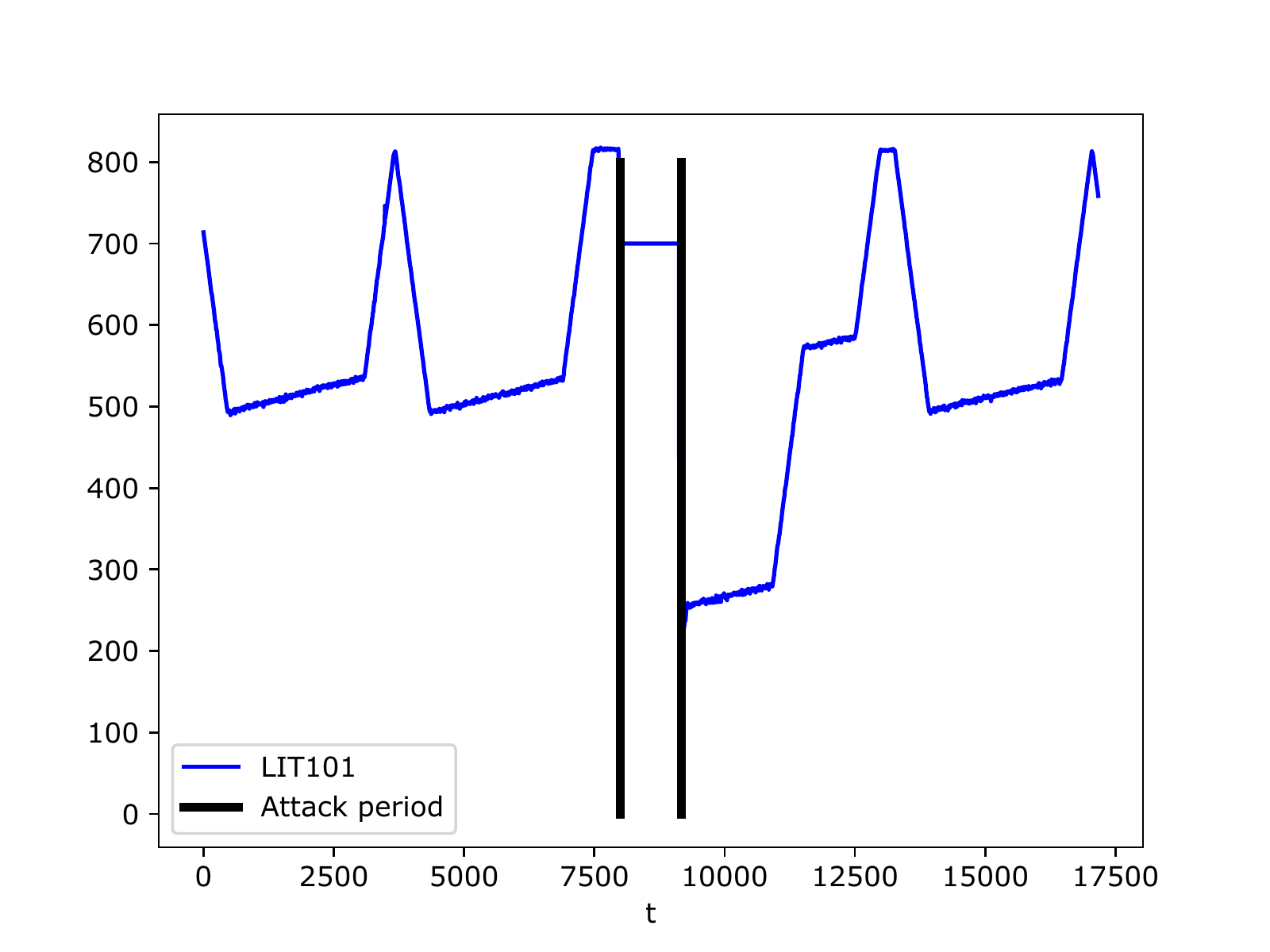}
\caption{Attack 30 and its impact on the system}
\label{fig:attack_30}
\end{figure}

As expected, the network cannot precisely predict the behavior of the system, which is highly non-linear and dependent on many factors, some of which are environmental and others of which are dictated by the controller program logic.
Moreover, it was observed that specific features can be predicted better than others.
Therefore, in order to detect anomalies we have chosen the following statistical approach. 
We calculate the absolute difference between the expected $\hat{y_t}$ and observed $y_t$ values for each feature at time $t$.
$$
	\vec{e_t} = \left| \vec{y} - \vec{\hat{y}} \right| \eqno{(1)}
$$
$\mu_e$ and $\sigma_e$ (the mean and standard deviation of the prediction error) are calculated overall of the data. 
In test time, for each prediction, we calculate the absolute value of the difference between the prediction and the observation and normalize it using the mean and standard deviation for each feature. We effectively have a \textit{z}-score of the probability of the prediction error for each feature: 
$$
	\vec{z_{e_t}} = \frac{\left| \vec{e_t} - \mu _e\right| }{\sigma _e} \eqno{(2)}
$$
A threshold for the \textit{z}-scores is used for the anomaly/attack detection.
If the maximal value across all the features passes this threshold, we enter the anomaly detection state. 
$$
	\max \vec{z_{e_t}} > T \eqno{(3)}
$$
However, due to irregularities and abrupt state changes in some features (e.g. pump on/off state) such deviations can exist but do not signify a cyberattack event. 
In order to account for this behavior and reduce the number of false alarms the threshold $T$ must be maintained for at least a specified duration of time, i.e. a time window denoted by $W$. 
The threshold $T$ and minimal time window $W$ are hyperparameters for the algorithm and are empirically determined as will be described later. 
$T$ basically defines the confidence level in the result.  
The final decision for an anomaly $A_i$ at time $i$ is determined by the following equation: 
$$
	A_i = \displaystyle \prod_{t=i-W}^i \max \vec{z_{e_i}} > T \eqno{(4)}
$$
Thus, we detect windows of anomalous behavior by deviation of prediction error from the observed statistics beyond the selected threshold for at least the selected time window. 
If the detected anomaly intersects with the attack period extended by a constant extra time immediately following the attack, we consider the attack to be detected.
The reason for adding the extra time is the fact that the attacks usually leave an impact on the system and the system's anomalous behavior occasionally continues much longer that the attack does.
We measure the detection performance based on the number of the attacks detected successfully (attack-based scoring).
We have found this detection method superior to the one used in \cite{inoue2017anomaly}, where the authors referred to the number of correctly detected log records.
The best way to explain our choice is by considering an example.
Figure \ref{fig:attack_30} shows the water level reported by the LIT101 sensor during and after an attack targeting the first stage of the system and aiming to underflow the first water tank (attack number 30 in the SWaT dataset). 

In order to reach the attack goal, the water level reported by the sensor is fixed at 700mm and the pump that pumps the water out of the tank is turned on continuously.
The black vertical lines in the figure indicate the attack period.
The figure shows a recovery period following the actual attack.
During this recovery period the system is not behaving normally, and thus detecting this time period as anomalous is correct.
If the detection performance is measured by correctly detecting the log records, then the records from the recovery period classified as anomalous would be considered false positives, as there was no attack at this time.
We believe that the attack-based scoring reflects the dynamics of ICS systems under attacks better than its log record-based alternative.
Another reason for our choice is the observation that the dataset is imbalanced and the number of attack log records represents about 12\% of all of the log records; thus the log records-based approach is biased unless we can balance the dataset (which is not trivial in our case).
As in real systems the ratio of attack to normal log records is expected to be even smaller, and thus we consider using log records-based precision and recall metrics less accurate. 

An alternative approach to anomaly detection used in \cite{goh2017anomaly} is the CUSUM  algorithm \cite{page1954continuous}. CUSUM is a classical algorithm for change detection in time series data. 
In CUSUM, the high and low cumulative sums, $SH$ and $SL$ respectively are calculated using the following regressive equation:
$$
SH_0 = SL_0 = 0
$$
$$
SH_{n+1} = \max (0,{S_n + x+n + \omega })  \eqno{(5)}
$$
$$
SL_{n+1} = \min(0,{S_n + x+n + \omega })
$$
In order to decide whether an anomaly has occurred, the authors of \cite{goh2017anomaly} compare the $SH$ and $SL$ to the \textit{Upper Control Limit (UCL)} and \textit{Lower control Limit (LCL)} correspondingly.
If $SH > UCL$ or $SL < LCL$ an anomaly is detected. 
The UCL and LCL were defined empirically for each feature. 
This algorithm was tested in our research as well but we did not use it for the following reasons:
\begin{itemize}
\item the statistical method produced better results, and 
\item the proposed CUSUM method requires choosing more hyperparameters (two per feature instead of just two).
\end{itemize}
\section{EXPERIMENT AND ANALYSIS}
\subsection{Setup}\label{ssec:setup}
Model training and testing was performed on two Intel i7-6700K workstations with 32GB of RAM using a NVIDIA 8GB 1080 GPU.
Training was performed until the validation loss stopped decreasing or until it hit its maximum iteration of 100 epochs. Training times varied greatly as we'll show in section \ref{ssec:results}. The models were implemented using Google's TensorFlow framework version 1.4 \cite{web:tensorflow}. Specifically, we used the low-level API which allowed for fine-grained control over the network architecture. 
We also evaluated an implementation based on the Keras library \cite{web:keras}, but found that  TensorFlow provided slightly faster training and test times. 
\subsection{Data Preprocessing}
\begin{figure}[t!]
\centering
\includegraphics[scale=0.5]{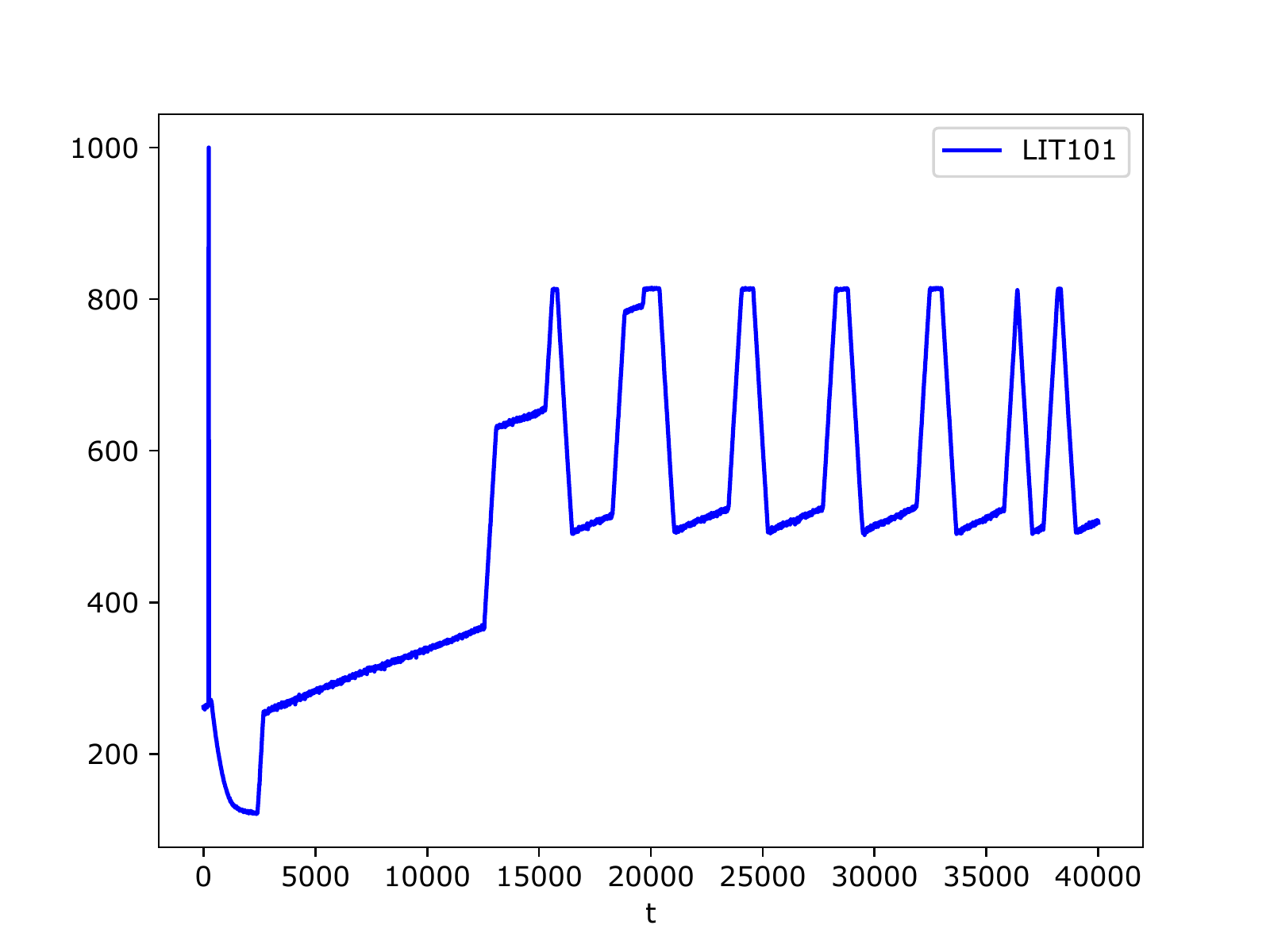}
\caption{Growth of the Water Level Measured by the LIT101 Sensor During the Initial Warm Up Period}
\label{fig:warm-up}
\end{figure}

Since this research focuses on physical layer attacks, we only used the "Physical" subdirectory containing the data collected from the sensors and actuators each second. 
The data collected under normal conditions has 496,800 data records and the data collected while performing attacks has 449,919 records with 36 attacks among them. The attacks usually span hundreds of seconds, with the shortest one being just 100 seconds and the longest one taking 35899 seconds, the second longest was 1,689 seconds.
The first 16,000 records of the training data were trimmed, as the system was unstable (see figure \ref{fig:warm-up}).
In addition, the data was normalized to (0,1) scale.
The minimal and maximal values of features from the training set were saved and used for scaling the test set.
It should be noted that in our final setup we scale both \textit{contiguous} (e.g. water level) and \textit{categorical} (e.g. motor valve closed) values to the (0,1) scale.
This generalizes our architecture to different data types.


Several data augmentation techniques were tested, but only one  produced an increase in detection score.
Specifically, in order to enrich the data with high-order features, the data is concatenated with newly engineered features representing the \textit{difference} between the current value of the feature and a past value with the given lag.
This difference carries the approximation of the derivative of the feature at the given moment of time.
$$
	\vec{x}_t =  \vec{x}_t | (\vec{x}_t - \vec{x}_{t-lag})\eqno{(7)}
$$
 
The final transformation we apply to the data is to divide it into batches to speed up the computations.
The number of batches we tried was varied from one to 100.
Larger numbers of batches provided faster training time.
The normal dataset (no attacks) was split: 80\% for training and 20\% for testing.

\begin{figure}[t!]
\centering
\includegraphics[clip, trim=0cm 9cm 0cm 0cm, width=\textwidth]{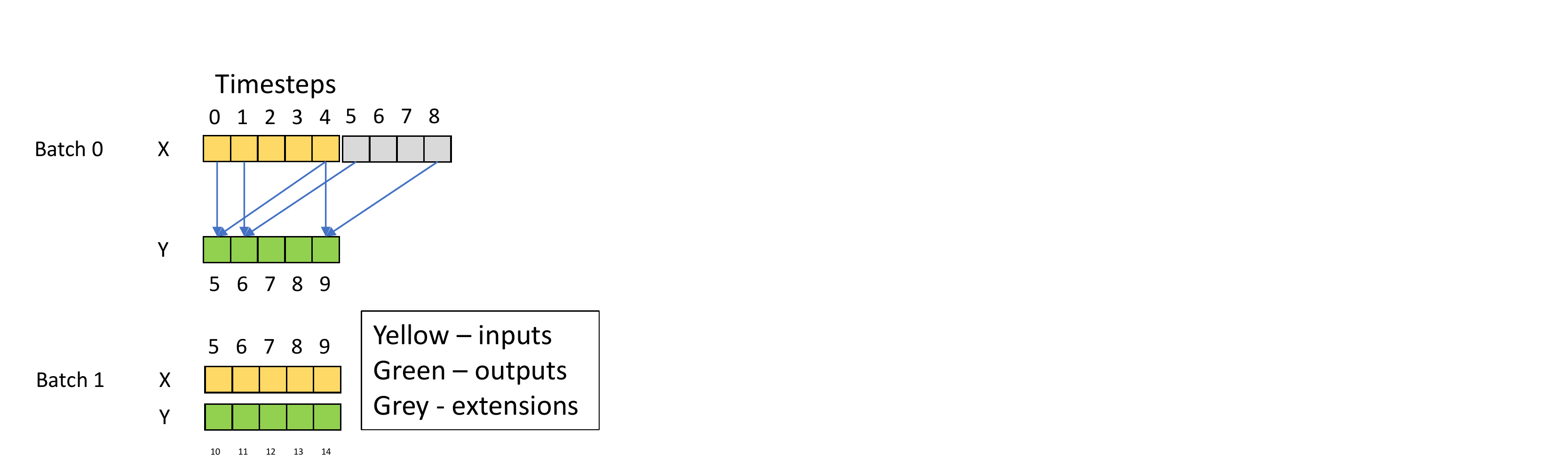}
\caption{Batches and their extension}
\label{fig:batches}
\end{figure}

Several prediction mode alternatives were tested.
One was a sequence-to-sequence model where $n$ timesteps of input were used to predict $m$ timesteps of output, where
$$
1 <= m <= n.\eqno{(8)}
$$
so that the sequence $(x_i, x_{i+1}, \ldots, x_{i+n-1})$ is used to predict a sequence $(y_j, y_{j+1}, \ldots, y_{j+m})$ where  $x_i$ and $y_j$ are input and output vectors of feature values at times $i$ and $j$ correspondingly. Also
$$
i < j < n + 1.\eqno{(9)}
$$
In other words, the predicted sequence is at most as long as the input one; the predicted sequence starts after the input one and either overlaps with it or strictly follows it.
Better results were achieved by predicting a single vector (i.e. sequence-to-vector prediction), so that
$$
m = 1
$$
$$
j = n.\eqno{(10)}
$$
The prediction mode described above necessitate that the data batches be extended in order to be able to predict all of the  output values.
This is best explained with an example.
Suppose there are 10 points split into two batches, each with five points.
Thus the first batch of five points is used to predict the first point in the second batch. 
Now consider the last four points of the second batch.
They are not predicted, as there are not enough points at the end of the first batch to predict them.
With 100 batches of 200 points each, there will be 19,900 points that the model haven't been trained on.
In order to solve this problem, the first batch is extended, appending the first 4 points of the second batch to the end of the first batch so that they can be used them to predict points six to nine in our example see figure \ref{fig:batches}.

\subsection{Architectures Used}
\begin{figure}[b!]
\centering
\includegraphics[clip, trim=0cm 7cm 0cm 0.5cm, width=\textwidth]{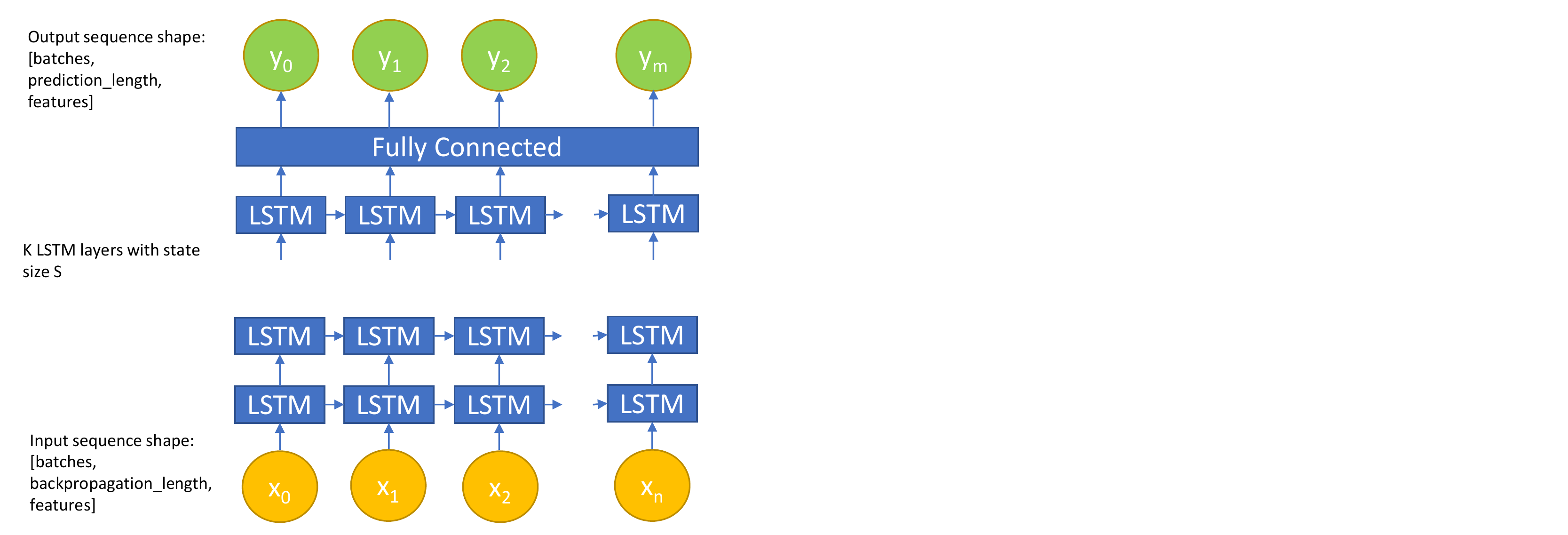}
\caption{Basic LSTM Model}
\label{fig:lstm}
\end{figure}

We began our experiments with recurrent neural networks, a popular neural network architecture approach for processing time series data\cite{lipton2015critical}.
As Figure \ref{fig:lstm} shows, we feed the network a sequence of length $N$, pass it through a number of stacked LSTM layers and apply a fully connected layer at the end to produce the prediction. 
Initially, we predicted the time window following the input, e.g. feeding the first 200 seconds of data in order to predict the following 200 seconds. 
This pattern continues, and in the next iteration we use seconds 201 - 400 to predict seconds 401 - 600 and so on. 
In later experiments we followed a different pattern, using only the last value of the output sequence as a prediction for the single data point immediately following the input sequence, i.e., as second 201 in our example.

\begin{figure}[t!]
\centering
\includegraphics[clip, trim=0cm 5.5cm 0cm 0cm, width=\textwidth]{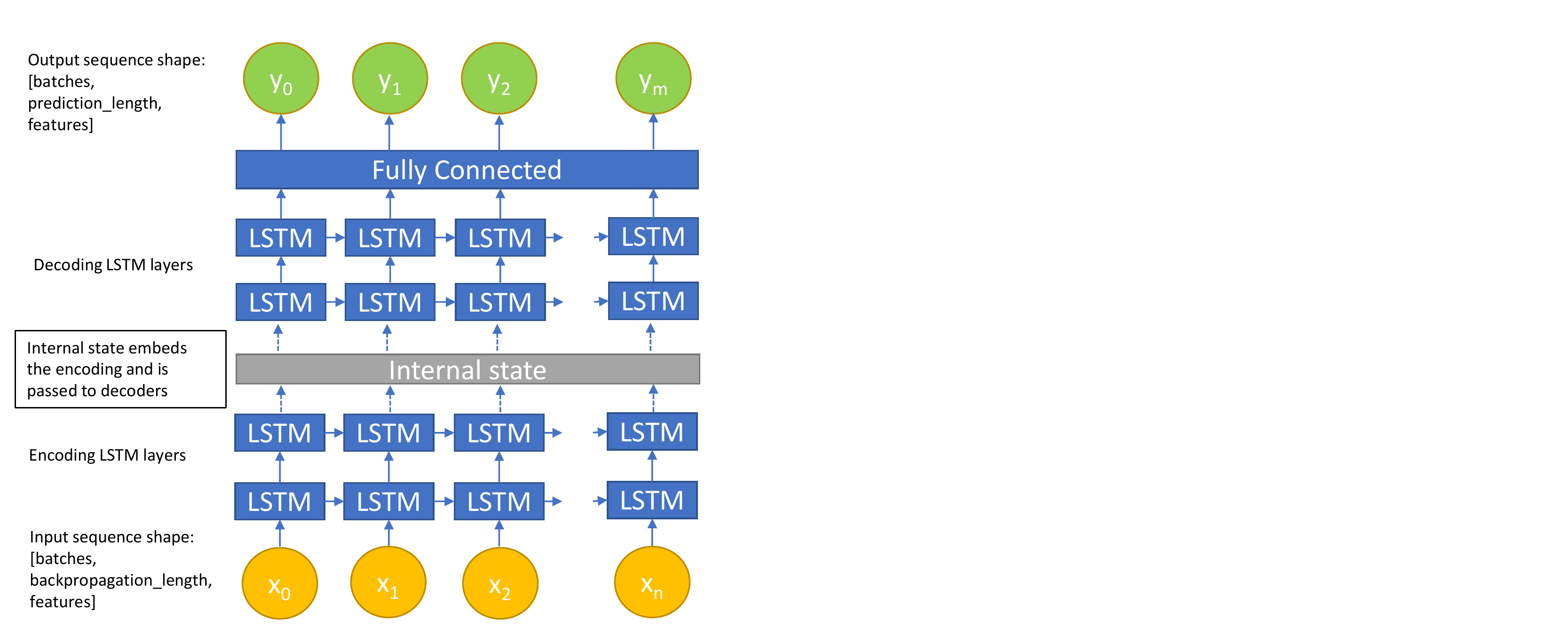}
\caption{LSTM Autoencoder Model}
\label{fig:lstm_autoenc}
\end{figure}

We use mean squared error (MSE) as a loss function and AdamOptimizer \cite{kingma2014adam} with learning decay for all experiments. 
$$
MSE = \frac{1}{n}\sum_{t=1}^{n}{Y_i - \hat{Y_i}} \eqno{(11)}
$$
We've experimented with multiple learning rates and learning decay strategies.
The learning rates were usually in the range of 0.001 - 0.00001, and the decay rate ranged from 0.9 to 0.99. 
We tested various depths of LSTM layers (from one to 10), state size range from 64 to 2048 and sequence lengths between 50 and 1000.
In addition to the basic LSTM recurrent network, we experimented with gated recurrent cell (GRU)\cite{cho2014learning}, replacing the LSTM cells and with bidirectional recurrent neural network architecture \cite{schuster1997bidirectional} based on LSTM cells. 
Another recurrent network architecture tested was an encoder-decoder LSTM network, based on the ideas in \cite{DBLP:journals/corr/SrivastavaMS15} and \cite{cho2014learning}. 
We encode the input sequence using layers of LSTMs, then we use the \textit{state} of the last encoder layer as an input to a stack of LSTM decoders' layers that predict the output sequence, as can be seen in Figure \ref{fig:lstm_autoenc}.
It should be noted that only the embedding represented by the internal state of the encoders is copied to the decoders, not the encoders output.

\begin{figure*}
\centering
\includegraphics[clip, trim=0cm 3cm 0cm 1.5cm, width=\textwidth]{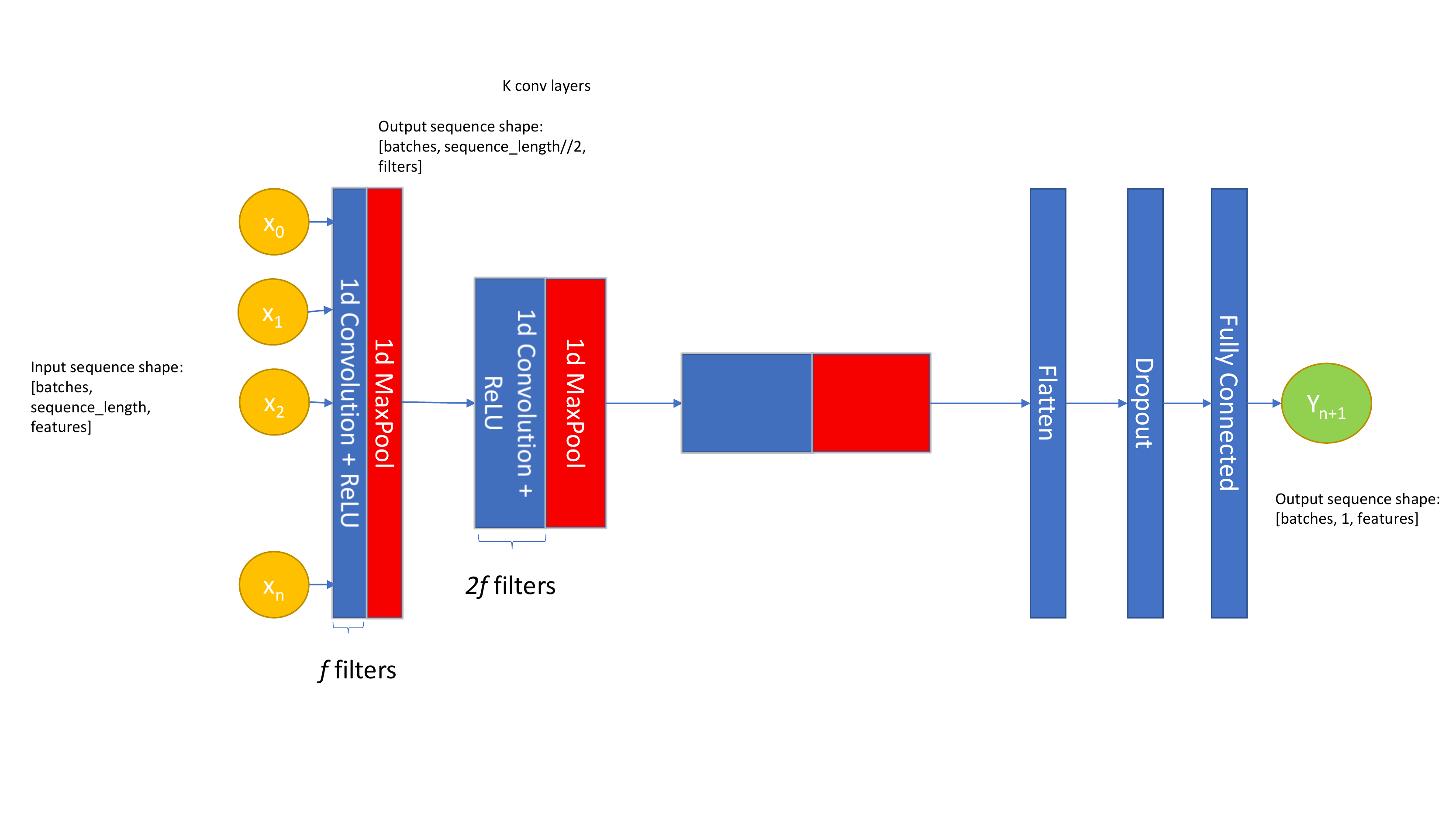}
\caption{1d Convolutional Neural Network Model}
\label{fig:cnn}
\end{figure*}

We also tested was convolutional neural networks, and more specifically 1D CNNs, which can be seen in Figure \ref{fig:cnn} in a number of alternative architectures.

\begin{figure}[b!]
\centering
\includegraphics[clip, trim=0cm 4cm 0cm 2.5cm, width=\textwidth]{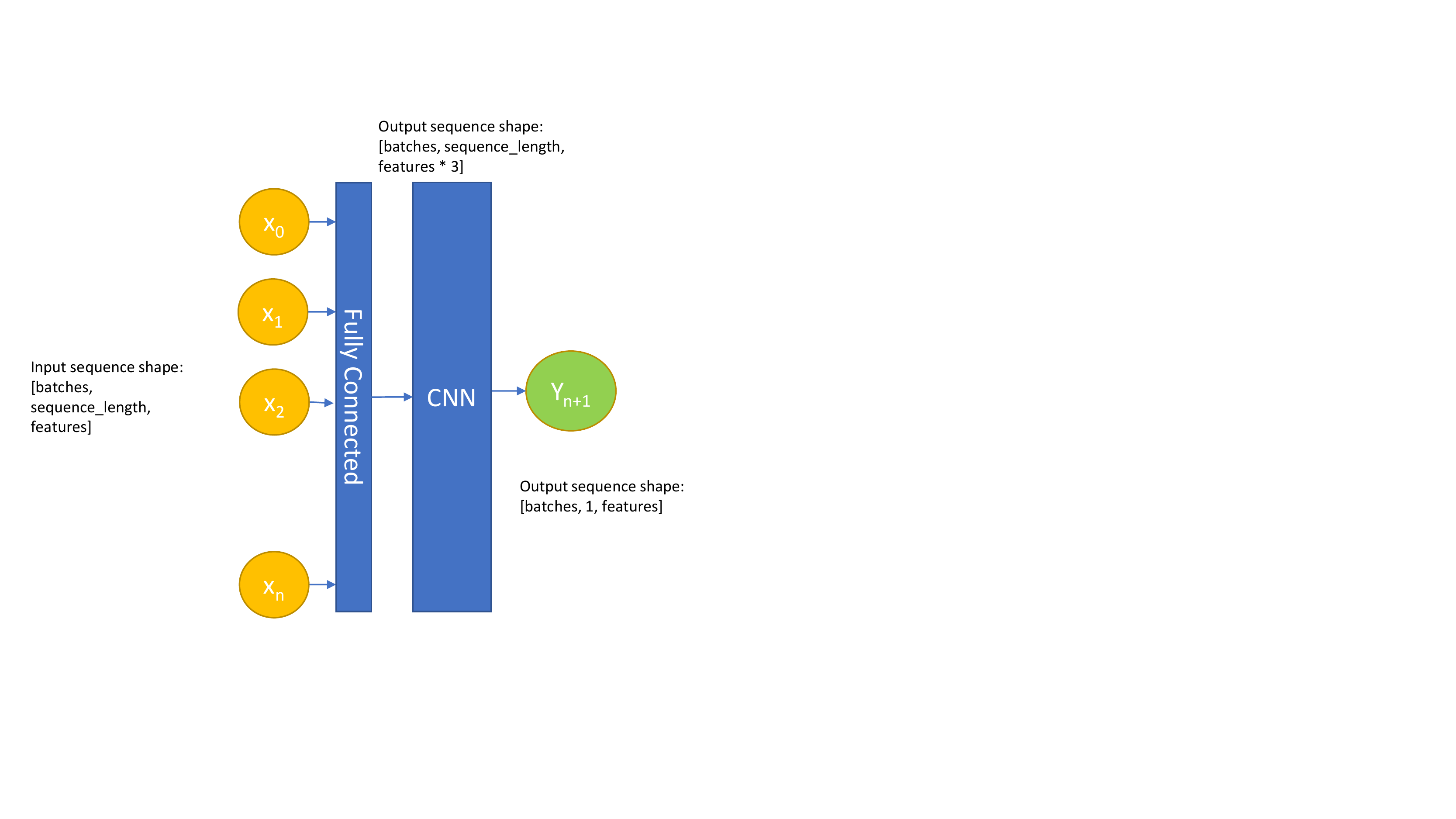}
\caption{Features Enrichment Before CNN}
\label{fig:fccnn}
\end{figure}

The most basic architecture follows the classical convolution-ReLU-MaxPooling scheme, where convolutions are 1D and applied to each feature separately along the time axis. 
We experimented with kernel sizes ranging from two to four and different filter sizes. 
Typically, we doubled filter size with each layer.
The stack of convolutional layers is followed by a fully connected layer that predicts the output. 
We used a dropout \cite{Srivastava:2014:DSW:2627435.2670313} layer to prevent overfitting.
As described above, the network is fed a sequence of points and predicts a single subsequent data point consisting of a number of features (see Figure \ref{fig:cnn}.
In an attempt to improve the performance of the network, we also experimented with batch normalization as described in \cite{Ioffe:2015:BNA:3045118.3045167}.
The batch normalization layer goes between the convolutional and ReLU layers.
As 1D convolutions are applied to single features along the time axis, the model only learns dependencies between combinations of features in the last fully connected layer. 
In order to infuse more knowledge about features' interdependencies we added an additional fully connected layer before the network, which extends the number of features we feed into the CNNs, as can be seen in Figure \ref{fig:fccnn}.

\begin{figure}[b!]
\centering
\includegraphics[clip, trim=0cm 4cm 0cm 0.5cm, width=\textwidth]{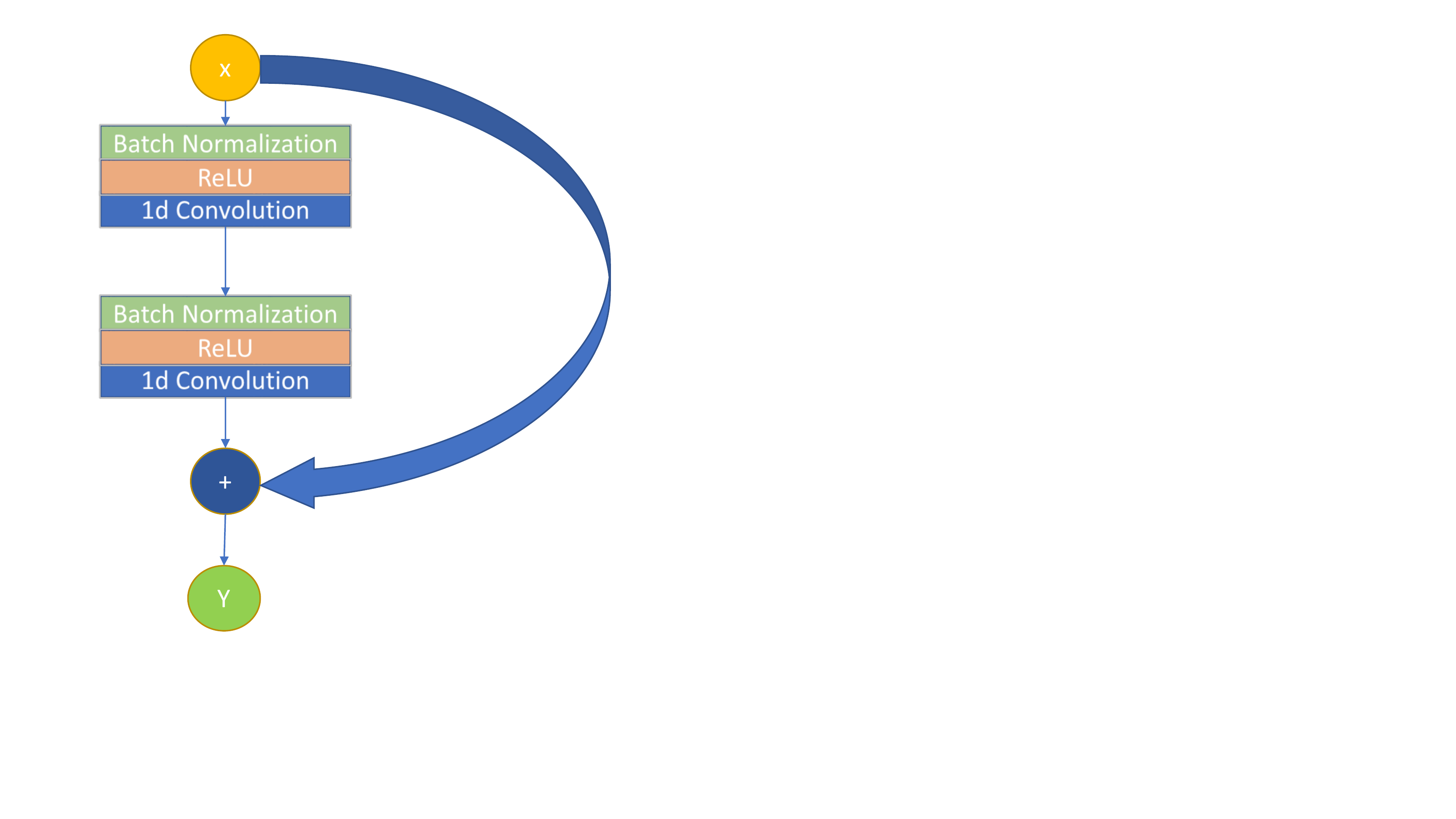}
\caption{ResNet block}
\label{fig:resnet}
\end{figure}

In addition to the classic CNN architecture (CONV-RELU-POOL), we tested other popular CNN architectures.
Specifically, we replaced the basic CONV-RELU-POOL block with $(CONV-RELU)\times N - MAXPOOL$ as found in VGG \cite{DBLP:journals/corr/SimonyanZ14a}.
We also experimented with replacing the convolutional layers with Inception layers following the original architecture described in \cite{DBLP:journals/corr/SzegedyLJSRAEVR14}.
Inception layers are known to provide superior performance while keeping computational cost low.
We used the implementation in \cite{web:Inception_Tensorflow} as a reference for the inception layer.
One of the most recent advances in CNN architecture involves the use of residual networks or ResNet \cite{DBLP:journals/corr/HeZRS15}.
We used a reference implementation of the more recent version of ResNet \cite{DBLP:journals/corr/HeZR016}, which can be seen in Figure \ref{fig:resnet}.

\begin{figure}[t!]
\centering
\includegraphics[clip, trim=0cm 4cm 0cm 0.5cm, width=\textwidth, scale = 0.8]{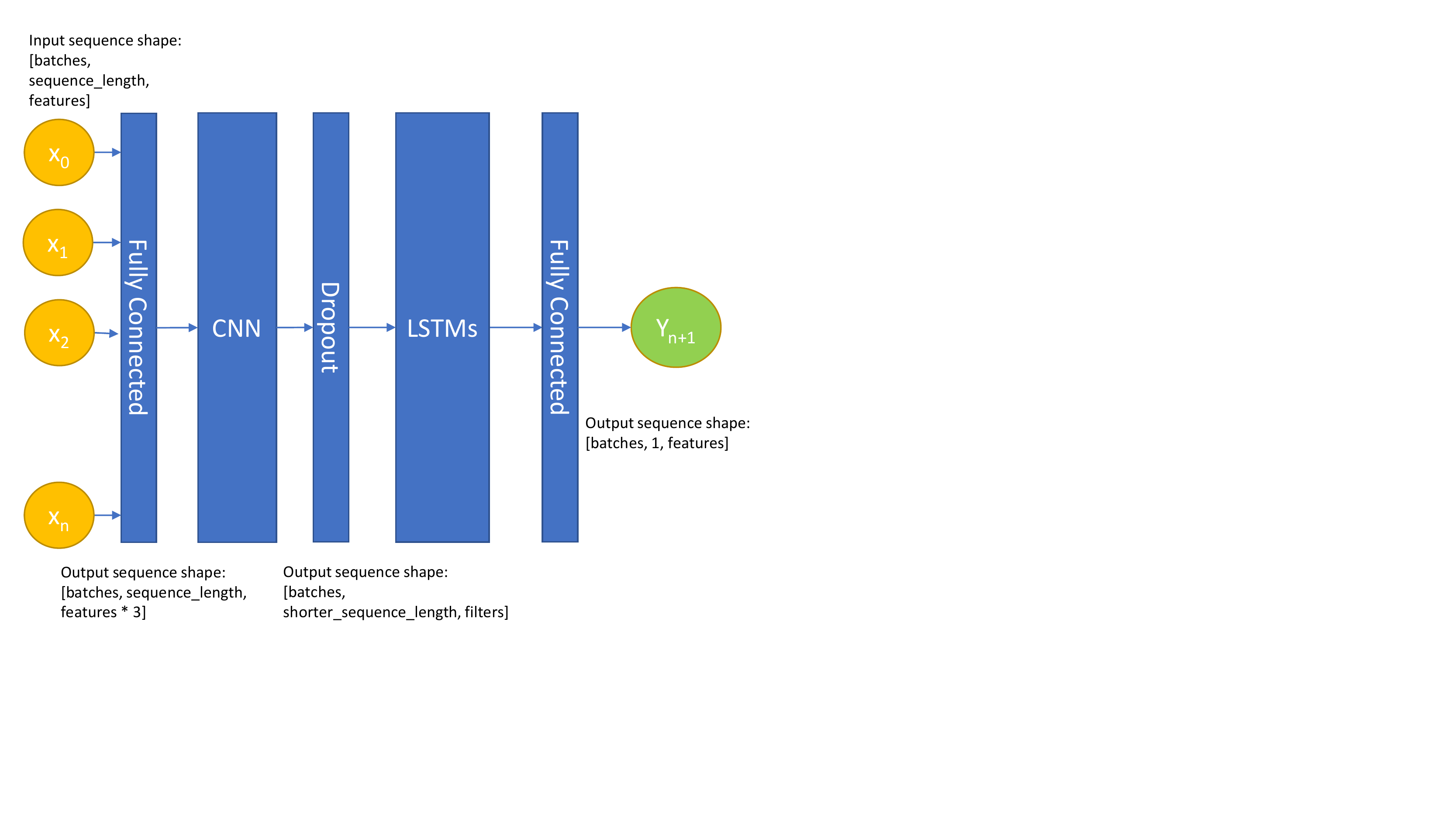}
\caption{Combined CNN + LSTM Model}
\label{fig:cnnlstm}
\end{figure}

Finally, we tested a combined architecture, where we process the data by a stack of convolutional (or inception) layers and then pass the output to LSTM layers, which make the prediction; this can be seen in figure \ref{fig:cnnlstm}.

\subsection{Hyperparameters Tuning}
First, we performed our tests on the data from the first stage (P1) of the six-stage process.
As P1 has only five features we were able to obtain the results relatively quickly. 
In addition, it should be easier to detect anomalies in the behavior of the first stage, as because at this point in the process as there are no prior stages influencing the internal system state.
Our tests had a large parameter space that included: input sequence length, network internal state size, output sequence length, the lag between input and output, network depth, dropout value and type (variational vs. regular), CNN kernel size and number of filters, pooling algorithm and size and others. 
We usually used a grid search strategy to explore the hyperparameters.
We selected $F1$ score as our performance metric.
$F1$ is calculated according to
$$
	F1=2\cdot{\frac{precision\cdot{recall}}{precision + recall}}\eqno{(11)}
$$
The precision and recall are calculated based on \textit{attack detection} rate according to the anomaly detection algorithm described in section \ref{ssec:anom_detect_method}. 
We chose $F1$ score of all possible $F_{\beta}$ scores, as we place equal importance on avoiding false positives and false negatives.
While conducting the experiments, we discovered that the detection results vary for multiple runs of the same configuration.
This is due to the low number of attacks and the resulting significant impact of a single mispredicted attack. 
Therefore we present average scores for multiple (usually five to 10) runs of the same configuration.
As explained in section \ref{ssec:anom_detect_method}, we use the neural network to predict the output and compare it to the actual output, detecting attacks using the window and threshold hyperparameters.
To determine the optimal hyperparameters for the model and stage we perform a grid search over time windows from 50 to 300 seconds and thresholds between 1.8 and 3.0, optimizing for the best $F1$ with the highest threshold to insure higher confidence. 

\subsection{Results} \label{ssec:results}
We first show that the selected neural network can learn the system features and predict them with adequate precision. 
We measure the prediction accuracy by observing the steady decrease in the training and validation error values until a sufficiently low value is reached for each.
We found that given enough computational power all of the neural networks were able to achieve an RMSE in range of 0.02 or less after a small numbers of training epochs. 
By "enough computational power" we refer to the size of the internal state for LSTMs, the number of filters in CNNs, and the number of layers in both.
We found that this level of error can usually be achieved with two or more layers.
Increasing it further  does not greatly improve the model accuracy, although some architectures converged faster than others.

\begin{figure}[t!]
\centering
\includegraphics[clip, trim=0cm 0cm 0cm 0cm, scale = 0.5]{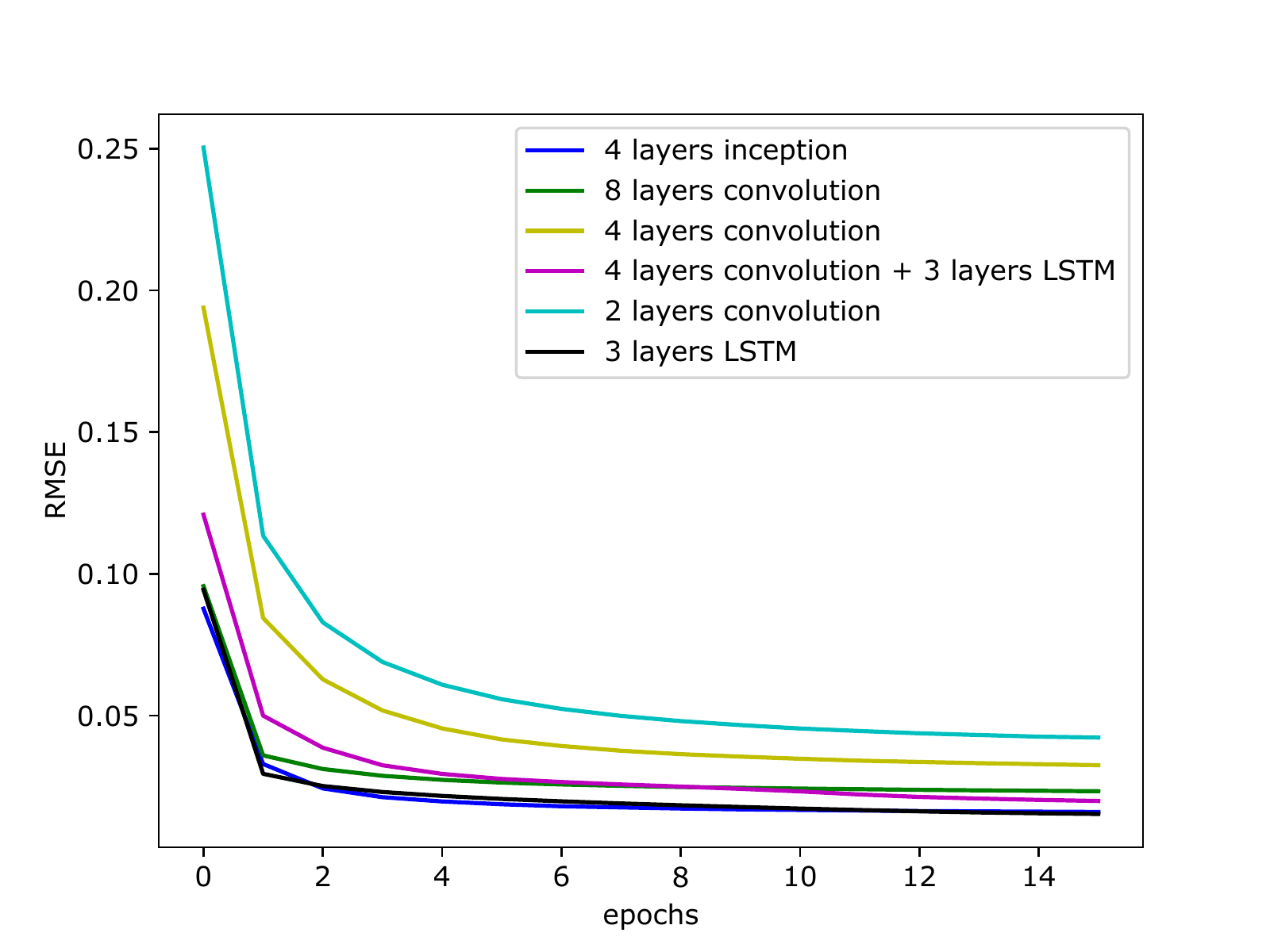}
\caption{Training Errors}
\label{fig:training_err}
\end{figure}
\begin{figure}[b!]
\centering
\includegraphics[clip, trim=0cm 0cm 0cm 0cm, scale = 0.5]{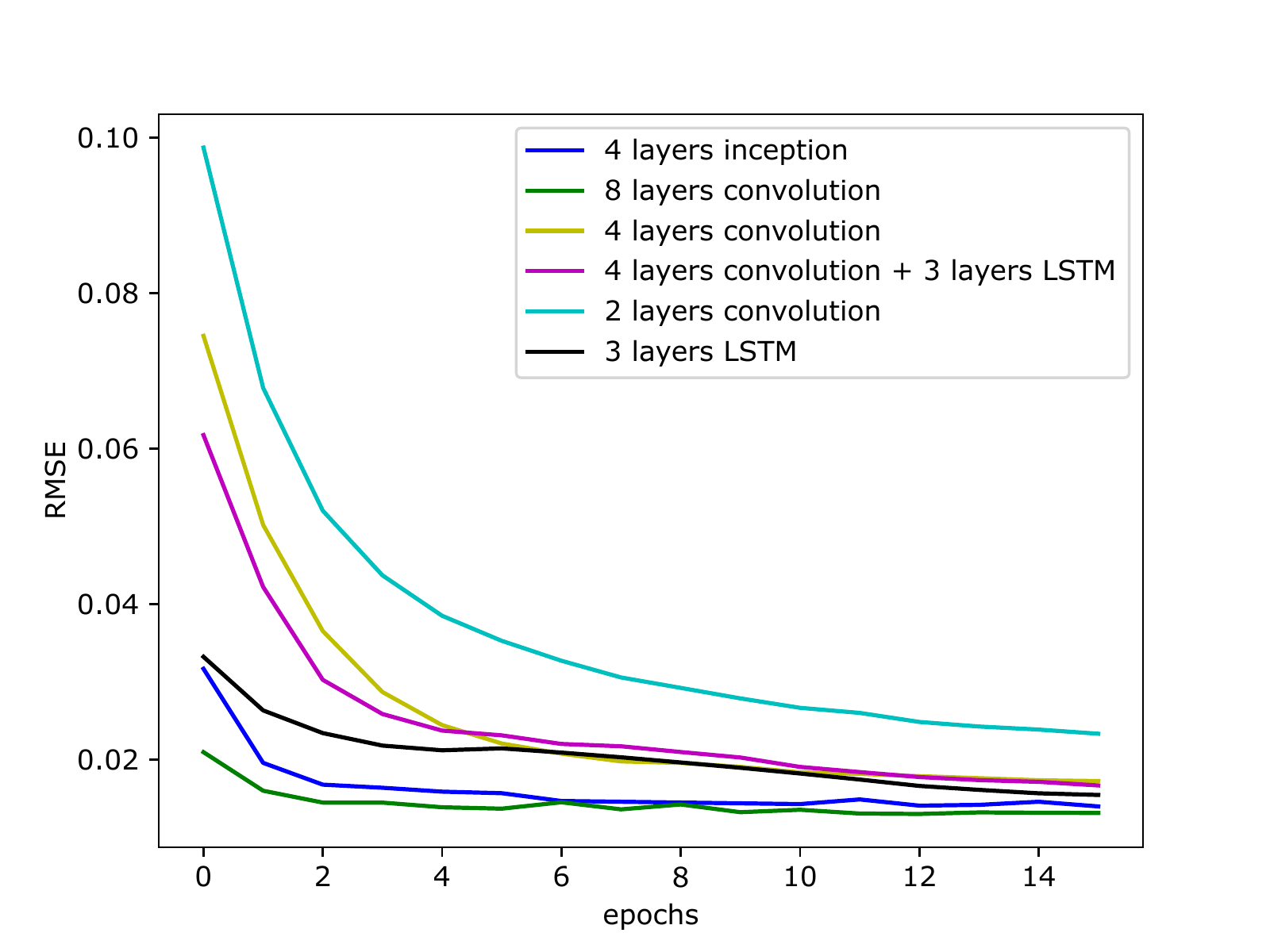}
\caption{Validation Errors}
\label{fig:val_err}
\end{figure}
\begin{figure}[t!]
\centering
\includegraphics[clip, trim=0cm 1cm 0cm 1cm, scale = 0.4]{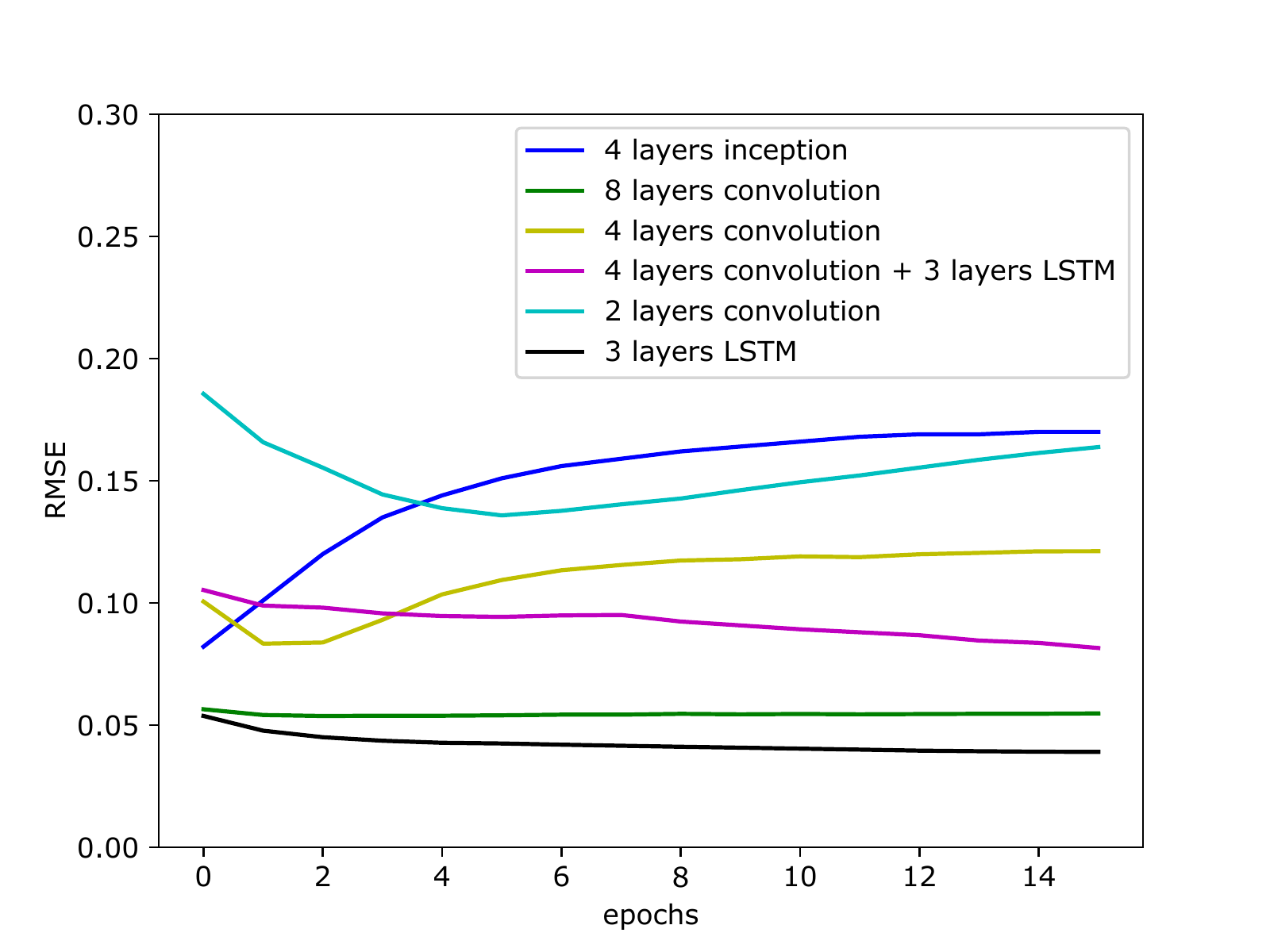}
\caption{Test Errors}
\label{fig:test_err}
\end{figure}

\begin{figure}[b!]
\centering
\includegraphics[clip, trim=0cm 0cm 0cm 1cm, scale = 0.45]{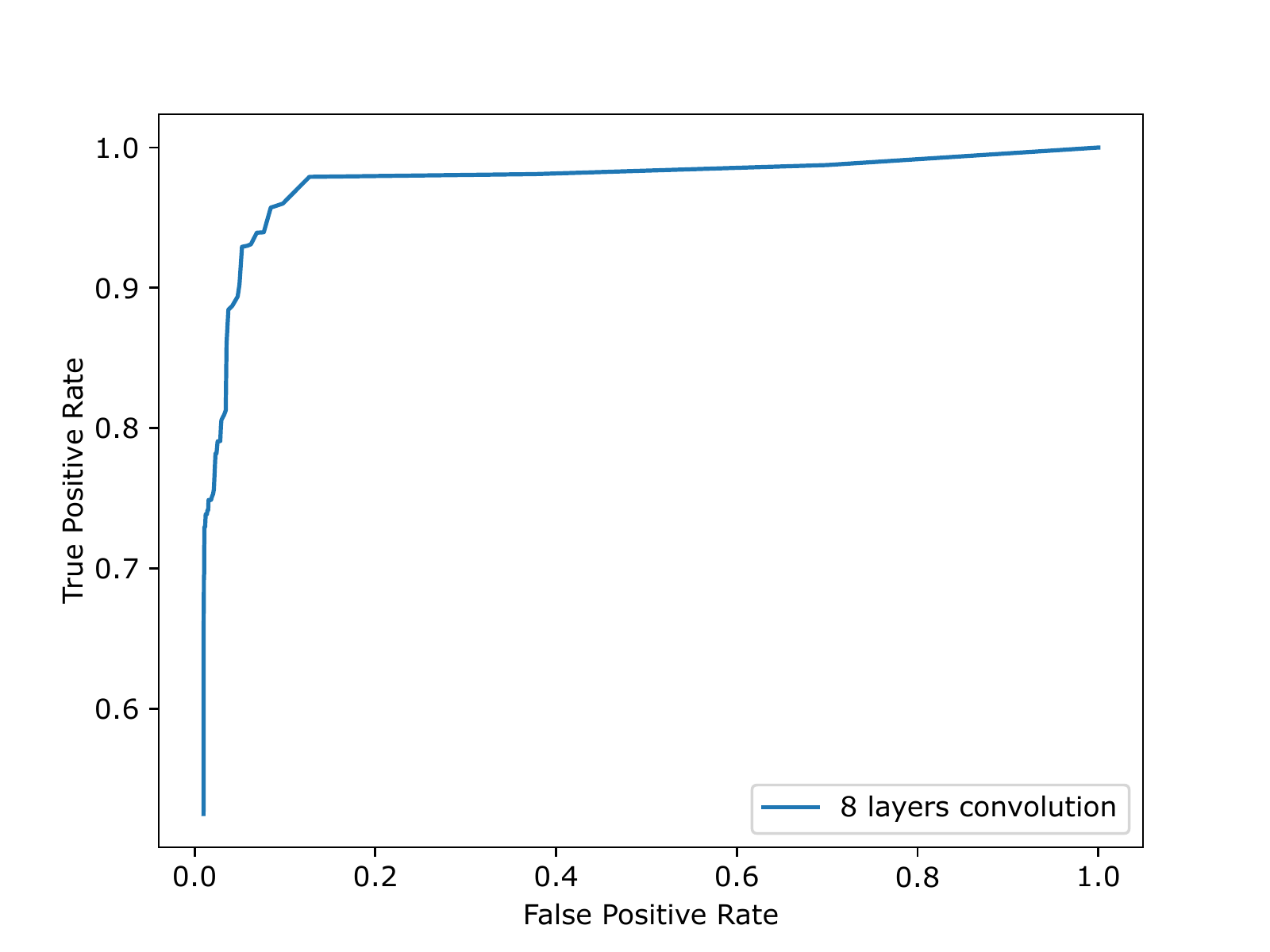}
\caption{ROC Curve For Eight Layers Convolutional Network}
\label{fig:ROC}
\end{figure}

As shown in Figure \ref{fig:training_err}, LSTMs and inception-based convolution converge the fastest and produce the lowest training error rate, but other configurations are not far behind and produce similar results with more iterations (only the first 16 iterations are shown).
As seen in Figure \ref{fig:val_err}, validation errors also consistently decreased, converging to the same error range.
The rate of test errors was also small, however LSTM-based models showed better convergence than pure CNNs, except for the eight layer CNN, which had a stable test error level.
We also estimated the AUC of the selected anomaly detection method based on per record detection metrics.
The reason for using per record labels for the AUC is the ease of defining the number of negative instances, which is problematic in the case of per-attack detection (it is impossible to measure how many non-attacks there are).
As Figure \ref{fig:ROC} shows, the anomaly detection algorithm provides high AUC, reaching 0.967 for the eight layers convolutional network.

\begin{figure}[t!]
\centering
\includegraphics[clip, trim=0cm 1cm 0cm 1cm, scale = 0.4]{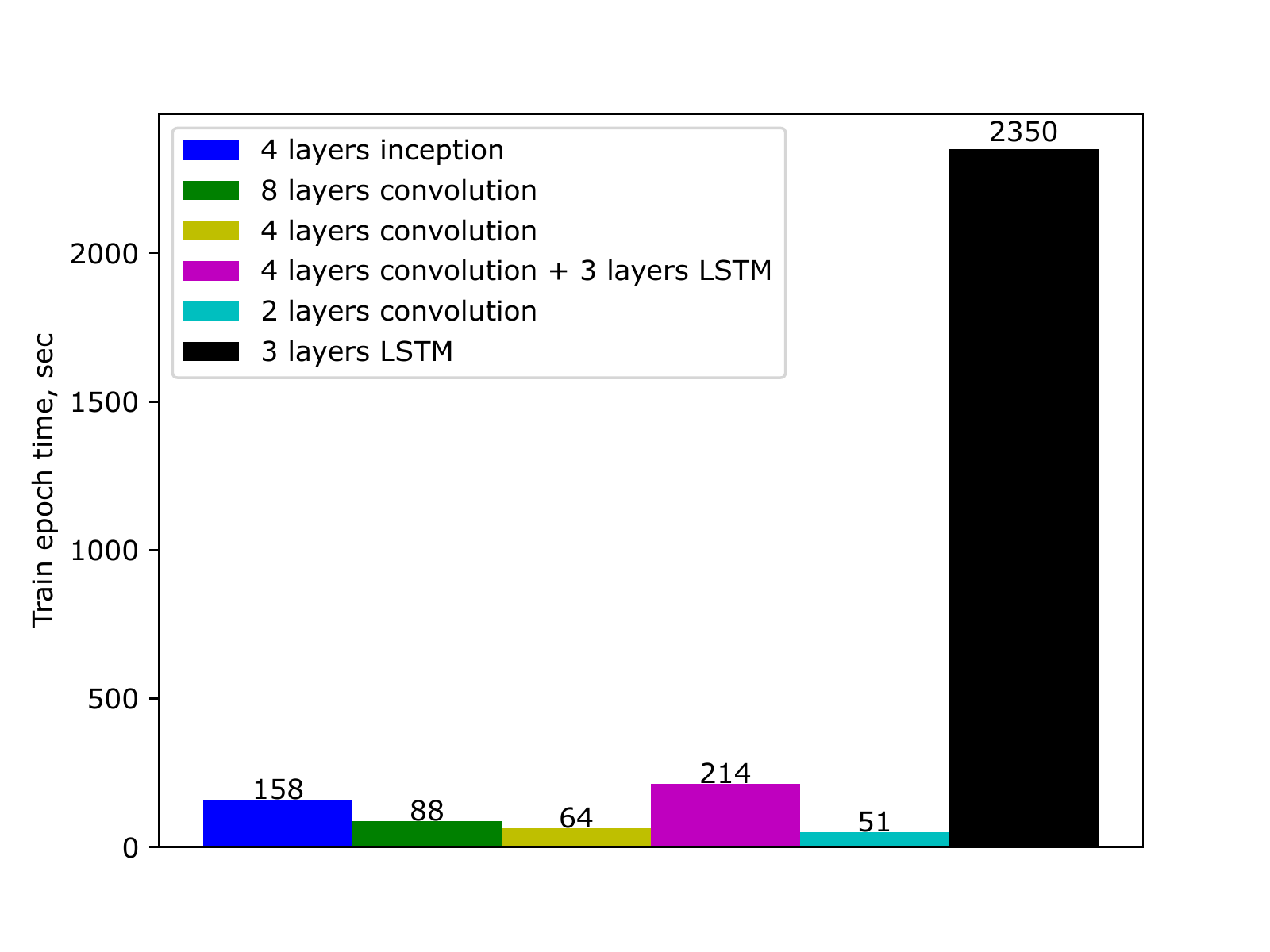}
\caption{Training times (One Epoch)}
\label{fig:train_time}
\end{figure}

\begin{figure}[t!]
\centering
\includegraphics[clip, trim=0cm 1cm 0cm 1cm, scale = 0.4]{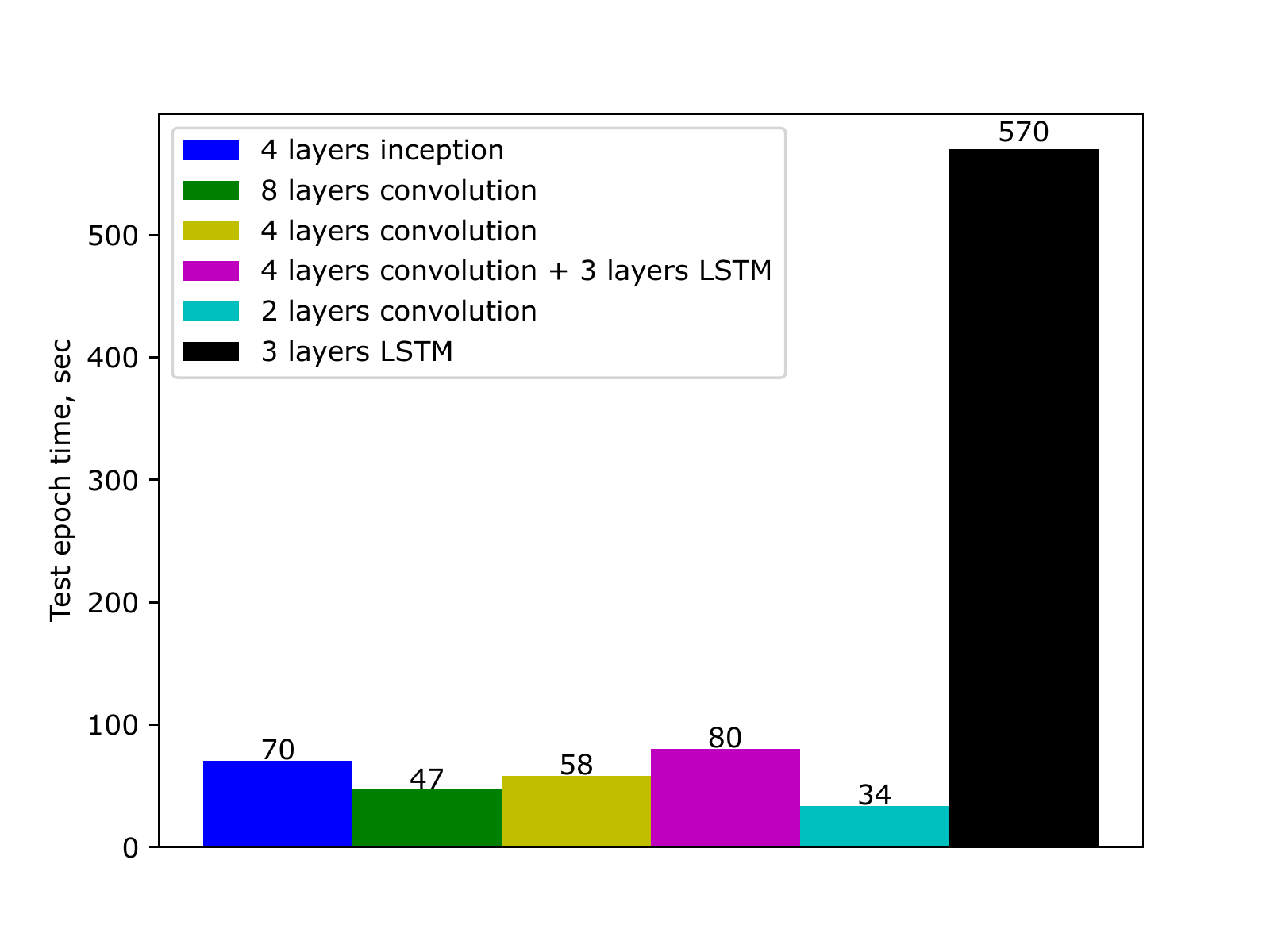}
\caption{Testing times (One Epoch)}
\label{fig:test_time}
\end{figure}

\begin{figure}[t!]
\centering
\includegraphics[clip, trim=0cm 1cm 0cm 1cm, scale = 0.4]{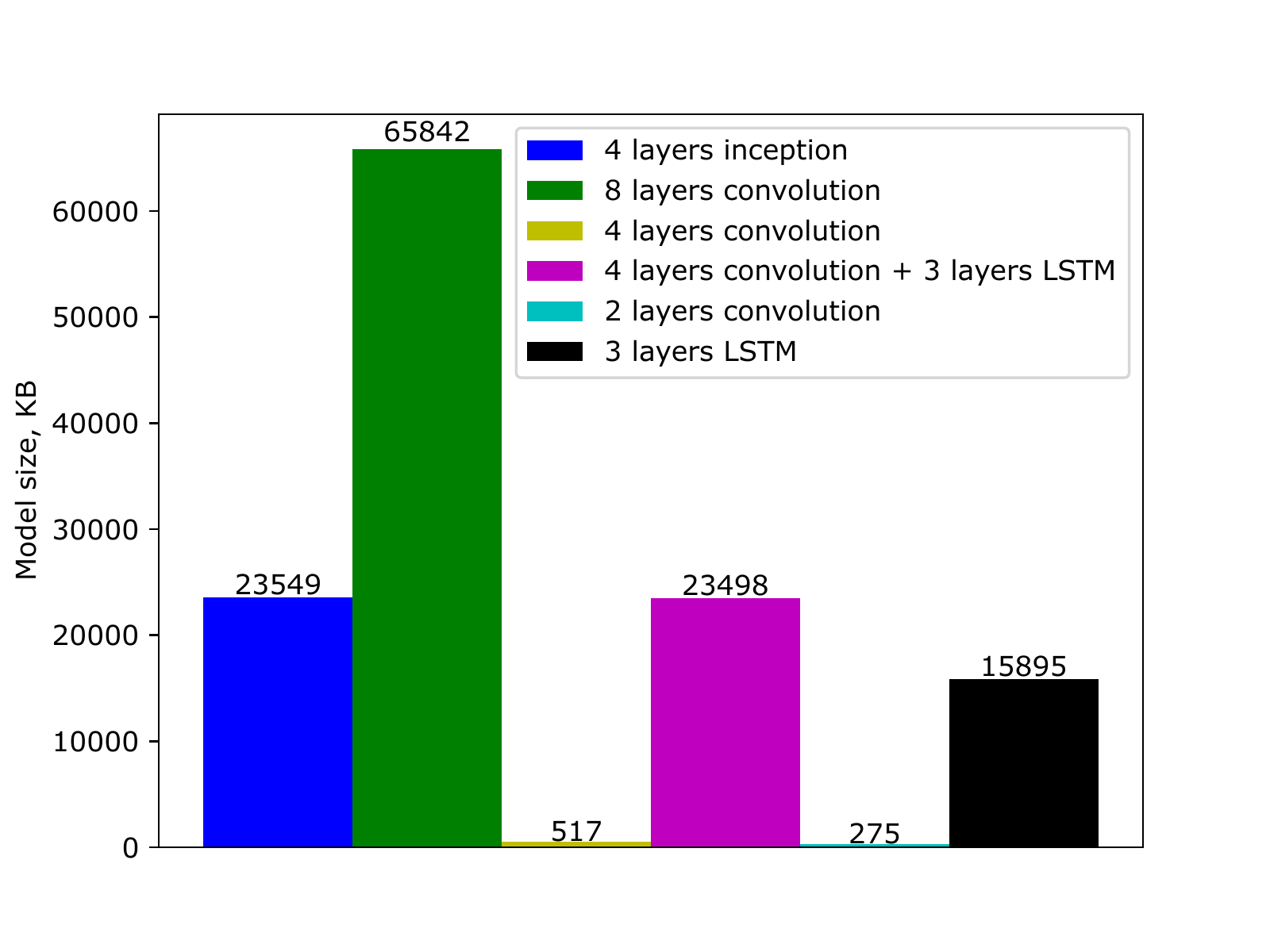}
\caption{Model Size}
\label{fig:model_size}
\end{figure}

It is interesting to compare the training and test times, as well as different model sizes, as can be seen in figures \ref{fig:train_time} - \ref{fig:model_size}.
The figures show average times per epoch, as measured at the workstation machines described in section \ref{ssec:setup}.
The training and testing times of convolutional networks were shorter by a factor 10 to 20 for testing and 15 to 40 for training compared to a pure LSTM network.
The training and testing times of a convolutional network with additional LSTM layers were also significantly shorter than of a pure LSTM network, as the convolutional layers significantly decreased the length of sequence input for the recurrent layers. 
As for the model sizes, adding more convolutional layers leads to significantly larger models, given our decision to double the number of filters in each layer.
These large models were still very fast to train and test.
Comparing the four layers CNN with a three layer LSTM shows that the LSTM model is 30 times larger than the CNN model.

\begin{figure}[t!]
\centering
\includegraphics[clip, trim=0cm 0cm 0cm 1cm, scale = 0.4]{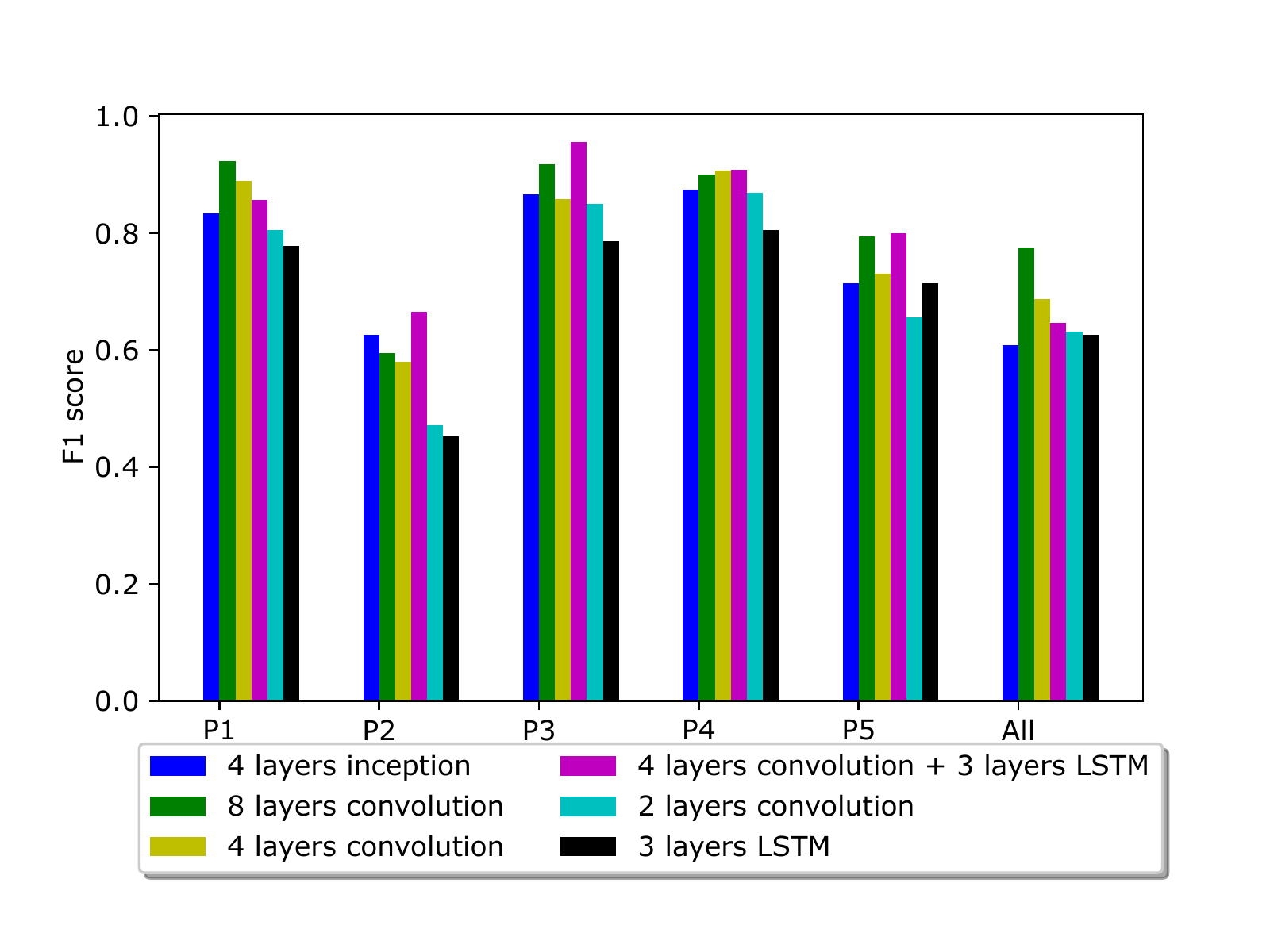}
\caption{F1 Scores Per Stage}
\label{fig:f1s}
\end{figure}
Next we present the attack detection performance comparison at individual stages.
The results presented are averaged over 5-10 runs of the same configuration.
As there are many hyperparameters for each configuration we present the best combination.
As Figure \ref{fig:f1s} shows, pure CNN networks demonstrated better anomaly detection results than their LSTM alternatives. 
We can also see that the inception layers had no advantage over simple convolutions.
The best results were achieved with eight layers convolutional network, but a four layers network was not far behind and sometimes surpassed the eight layers CNN, while being much smaller and faster to train.
There are two stages in the SWaT testbed, P2 and P6, which achieved inferior results (average $F1 < 0.6$). 
In order to understand the reason for this, one needs to consider the nature of attacks related to these stages.
For example, in the P6 stage, some of the sensors and actuators were not used for data collection; in addition, there is only one attack that involves this stage, thus its detection results are not representative.
Other works using this dataset \cite{lin2018tabor} also excluded P6 from the study.
As for the attacks in the P2 satage, there are only four (6,24,29,30) such attacks in the database and they include one (29) that did not achieve the desired effect (due to a mechanical interlock) and  another (24) that did not have the expected impact according to the test description.
Thus we can conclude that the relatively low $F1$ score for the P2 stage is caused by the test setup and our model successfully detected the attacks were actually carried out.

In order to determine which attacks are not discovered by our detection method (the tests were done on a four and eight layer convolutional networks), we took an ensemble approach, and combined the detection results for individual stages. 
We discovered that only four attacks were consistently undetected: attack numbers 4, 13, 14 and 29. 
When reviewing the description of these tests we found that these attacks failed to have their expected impact on the system; thus, it is reasonable that they were undetected.
When combining the attacks detected in individual stages, the model successfully detected 32 of 36 attacks, thus improving upon the reported results obtained in \cite{inoue2017anomaly}.
The $F1$ score of the ensemble of four layers 1D CNN model was 0.9206, with precision of 1 and recall of 0.8529. 
We also observed that when trained on the data of all of the the stages together, the model achieves slightly lower $F1$: 0.688 for four layers and 0.775 for eight layers.
In contrast to performing per attack calculations, when calculating per log record detection, the $F1$ for the model trained on the data of all of the stages together was 0.817 for four layers and 0.861 for eight layers, thus improving over previously shown results \cite{inoue2017anomaly}.
We estimate that the reason the $F1$ of the single combined model is smaller than the ensemble $F1$ of the individual stages is that the proposed architectures were too small to capture all the complex dependencies across different stages including time delays between these dependencies.
Tables \ref{table:results} and \ref{table:scores} summarize the averaged results of the best working network configuration presented in the figures above.
The main hyperparameters of the configurations appear in the tables.
While this paper was being written, new research was published \cite{lin2018tabor}, which uses a novel graphical model-based approach for anomaly detection with the SWaT data set.
In order to make a comprehensive comparison with previous work, a recall evaluation per attack scenario was conducted and the results of this evaluation are presented in tables \ref{table:perf_comparison} and \ref{table:attack_recall}.
The comparison shows that a 1D CNN provides recall improvement in almost all of the cases.
It should be noted that in table \ref{table:attack_recall} the recall was calculated based on attack detection and not on individual record detection as explained earlier in section \ref{ssec:anom_detect_method}.

\begin{table*}[h]
\caption{Test Results} \label{table:results}
\centering
\begin{tabular}{|c|c|c|c|}
\hline
Network configuration & Train epoch time, sec & Test epoch time, sec & Model Size, KB\\
\hline
4 inception layers, kernel = 2, 32 filters & 158.69 & 70.8 & 23549\\
\hline
8 convolutional layers, kernel = 2, 32 filters & 88	& 47 & 65842\\
\hline
4 convolutional layers, kernel = 2, 32 filters & 64 & 58 & 517\\
\hline
4 convolutional layers + 3 LSTM layers, kernel = 2, 32 filters & 214 & 80 & 23498\\
\hline
2 convolutional layers, kernel = 2, 32 filters & 51 & 34 & 275\\
\hline
3 LSTM layers with state = 256 & 2350 & 570 & 15895\\
\hline
\end{tabular}
\end{table*}
\begin{table*}[]
\caption{$F1$ Scores Per Stage} \label{table:scores}
\centering
\begin{tabular}{|c|c|c|c|c|c|c|c|}
\hline
Network configuration & P1 & P2 & P3 & P4 & P5 & All\\
\hline
4 inception layers, kernel = 2, 32 filters & 0.834 & 0.626 & 0.866 & 0.875 & 0.714 & 0.609\\
\hline
8 convolutional layers, kernel = 2, 32 filters & 0.924 & 0.595 & 0.918 & 0.901	& 0.795 & 0.775\\
\hline
4 convolutional layers, kernel = 2, 32 filters & 0.89 & 0.58 & 0.858 & 0.907 & 0.731
 & 0.688\\
\hline
4 convolutional layers + 3 LSTM layers, kernel = 2, 32 filters & 0.857 & 0.666 & 0.956 & 0.909 & 0.8 & 0.646\\
\hline
2 convolutional layers, kernel = 2, 32 filters & 0.805 & 0.472 & 0.85 & 0.87 & 0.656 & 0.632
 \\
\hline
3 LSTM layers with state = 256 & 0.778 & 0.453 & 0.787 & 0.805 & 0.714 & 0.626\\
\hline
\end{tabular}
\end{table*}

\begin{table*}
\caption{Attack detection performance comparison} \label{table:perf_comparison}
\centering
\begin{tabular}{|c|c|c|c|}
\hline
Method & Precision & Recall & F1\\
\hline
DNN & 0.98295 & 0.67847 & 0.80281\\
\hline
SVM & 0.92500 & 0.69901 & 0.79628\\
\hline
TABOR & 0.86171 & 0.78803 & 0.82322\\
\hline
1D CNN combined records & 0.95 & 0.656 & 0.766\\
\hline
1D CNN combined attacks & 0.95 & 0.787 & 0.861\\
\hline
1D CNN ensembled records & 0.867 & 0.854 & 0.860\\
\hline
1D CNN ensembled attacks & 1.0 & 0.853 & 0.920\\
\hline
\end{tabular}
\end{table*}

\begin{table*}
\caption{Point recall evaluation in each attack scenario for 4 layers 1D CNN.} \label{table:attack_recall}
\centering
\begin{tabular}{|c|c|c|c|c|c|c|}
\hline
No. & No. Scenario & Description Of Attack & DNN & SVN & TABOR & 1D CNN\\
\hline
1 & 1 & Open MV-101 & 0 & 0 & 0.049 & 1\\
\hline
2 & 2 & Turn on P-102 & 0 & 0 & 0.930 & 1\\
\hline
3 & 3 & Increase LIT-101 by 1mm every second & 0 & 0 & 0 & 0.9\\
\hline
4 & 4 & Open MV-504 & 0 & 0.035 & 0.328 & 0\\
\hline
5 & 6 & Set value of AIT-202 as 6 & 0.717 & 0.720 & 0.995 & 1\\
\hline
6 & 7 & Water level LIT-301 increased above HH & 0 & 0.888 & 0 & 1\\
\hline
7 & 8 & Set value of DPIT as \textless40kpa & 0.927 & 0.919 & 0.612 & 1\\
\hline
8 & 10 & Set value of FIT-401 as \textgreater0.7 & 1 & 0.433 & 0.994 & 1\\
\hline
9 & 11 & Set value of FIT-401 as 0 & 0.978 & 1 & 0.998 & 1\\
\hline
10 & 13 & Close MV-304 & 0 & 0 & 0 & 0\\
\hline
11 & 14 & Do not let MV-303 open & 0 & 0 & 0 & 0\\
\hline
12 & 16 & Decrease water level LIT-301 by 1mm each second & 0 & 0 & 0 & 1\\
\hline
13 & 17 & Do not let MV-303 open & 0 & 0 & 0.597 & 1\\
\hline
14 & 19 & Set value of AIT-504 to 16 uS/cm & 0.123 & 0.13 & 0.004 & 1\\
\hline
15 & 20 & Set value of AIT-504 to 255 uS/cm & 0.845 & 0.848 & 0.997 & 1\\
\hline
16 & 21 & Keep MV-101 on continuously; value of LIT-101 set as 700mm & 0 & 0.0167 & 0.083 & 1\\
\hline
17 & 22 & Stop UV-401; value of AIT502 set as 150; force P-501 to remain on & 0.998 & 1 & 0.998 & 1\\
\hline
18 & 23 & Value of DPIT-301 set to \textgreater0.4 bar; keep MV-302 open; keep P-602 closed & 0.867 & 0.875 & 0 & 1\\
\hline
19 & 24 & Turn off P-203 and P-205 & 0 & 0 & 0 & 1\\
\hline
20 & 25 & Set value of LIT-401 as 1000; P402 is kept on & 0 & 0.009 & 0 & 0.2\\
\hline
21 & 26 & P-101 is turned on continuously; set value of LIT-301 as 801mm & 0 & 0 & 0.999 & 1\\
\hline
22 & 27 & Keep P-302 on continuously; value of LIT401 set as 600mm till 1:26:01 & 0 & 0 & 0.196 & 1\\
\hline
23 & 28 & Close P-302 & 0.936 & 0.936 & 1.000 & 1\\
\hline
24 & 29 & Turn on P-201; turn on P-203; turn on P-205 & 0 & 0 & 0 & 0\\
\hline
25 & 30 & \vbox{\hbox{\strut Turn P-101 on continuously; turn MV-101 on continuously;set value of}\hbox{\strut LIT-101 as 700mm; P-102 started itself because LIT301 level became low}} & 0 & 0.003 & 0.999 & 1\\
\hline
26 & 31 & Set LIT-401 to less than L & 0 & 0 & 0 & 0.2\\
\hline
27 & 32 & Set LIT-301 to above HH & 0 & 0.905 & 0 & 1\\
\hline
28 & 33 & Set LIT-101 to above H & 0 & 0 & 0.890 & 1\\
\hline
29 & 34 & Turn P-101 off & 0 & 0 & 0.990 & 1\\
\hline
30 & 35 & Turn P-101 off; keep P-102 off & 0 & 0 & 0.258 & 0.2\\
\hline
31 & 36 & Set LIT-101 to less than LL & 0 & 0.119 & 0.889 & 1\\
\hline
32 & 37 & Close P-501; set value of FIT-502 to 1.29 at 11:18:36 & 1 & 1 & 0.998 & 1\\
\hline
33 & 38 & Set value of AIT402 as 260; set value of AIT502 to 260 & 0.923 & 0.927 & 0.996 & 1\\
\hline
34 & 39 & Set value of FIT-401 as 0.5; set value of AIT-502 as 140 mV & 0.940 & 0 & 0.369 & 1\\
\hline
35 & 40 & Set value of FIT-401 as 0 & 0.933 & 0.927 & 0.997 & 1\\
\hline
36 & 41 & Decrease LIT-301 value by 0.5mm per second & 0 & 0.357 & 0 & 1\\
\hline
\end{tabular}
\end{table*}

\section{CONCLUSIONS AND FUTURE WORK}
In this paper, we studied the use of deep learning architectures, namely convolutional and recurrent neural networks for cyberattacks detection in ICSs.
We tested a number of architectures and hyperparameters and proposed a statistical window-based anomaly detection method that has been shown effective in detecting the attacks in the SWaT dataset. 
We introduced a model based on 1D convolutional neural network that successfully detected 32 out of 36 attacks.
Its superior run time and performance have been elaborated upon and show great promise for ICS cyberattack detections.

In our proposed 1D CNN architecture we apply 1D convolutions to each feature across the time dimension.
This has a number of potential limitations.
First, the model only learns mutual dependencies between the features in the very last fully connected layer. 
In order to address this limitation, we introduced a fully connected layer in the very beginning of the data flow, enriching the feature space.
Second, 1D CNNs are stateless and lack the ability to learn beyond the sequence used as a sample.
We addressed this by augmenting the feature space by adding features' first derivatives as additional features. 
This provided slightly better results, however we found that using second derivatives and a few other methods did not result in additional improvements.

In the current research we found that having a dedicated model for each stage produces better results than having a single model for the whole system.
One possible reason for that relates to the limitations of the hardware used for building a model large enough to represent all of the stages and their dependencies (including time dependencies) together.
In addition to superior results, training separate models for stages will scale better than a single model, but the ways to learn inter-stage dependencies need to be examined.

The research performed was not based on assumptions regarding the modeled system and its attacks, and thus we expect it results to be generic.
At the same time, one potential threat to the research validity is the fact that it was tested on a single dataset from a single type of industrial process.
Another potential validity limiting direction is the fact that the attacks were relatively simple and spoofed only small subset of features.

As the research focused on the ability to detect cyberattacks, it would be beneficial to test the method's ability to detect faulty equipment behavior, which is another kind of anomalous ICS behavior.

Beyond addressing these limitations this research can be expanded in a number of directions:
\begin{itemize}
\item studying methods for learning cross-stage behavioral features,
\item studying methods for embedding high-order process modeling for anomaly detection,
\item investigating the application of recent audio generative network architectures, e.g. WaveNet \cite{van2016wavenet} to anomaly detection,
\item applying the proposed anomaly detection method to streaming data and comparing its performance other real-time algorithms,
\item improving the interpretability of the results, and
\item applying the anomaly detection method to faulty ICS equipment behavior detection.
\end{itemize}








\bibliographystyle{ACM-Reference-Format}
\bibliography{refs} 

\end{document}